\documentclass[showpacs,preprintnumbers,amsmath,amssymb,nofootinbib,superscriptaddress,twocolumn]{revtex4}

\usepackage{graphicx}% Include figure files
\usepackage{amsmath}
 \usepackage{color}
\usepackage{changepage}
\usepackage{subfigure}
\usepackage{multirow}
\usepackage{epstopdf}
\usepackage{booktabs}

\begin{document}

\title{Reverse-graph Stationary Probability Dependent Random Walks on Directed Graphs}
%{A Novel Random Walk Strategy on Directed Graphs}
%Optimizing Controlling the transport efficiency on Cayley trees Controlling the efficiency of trapping
%\shorttitle{} %

\author{Shanshan Luo}
\affiliation{
  School of Math and Information Science, Guangzhou University, Guangzhou 510006, China.}
\affiliation{
Guangdong Provincial Key Laboratory co-sponsored by province and city of Information Security Technology, Guangzhou University, Guangzhou 510006, China.}
\affiliation{
Guangzhou Center for Applied Mathematics, Guangzhou University, Guangzhou 510006, China.}

\author{Junhao Peng \thanks{Author to whom correspondence should be addressed.}}
\email{pengjh@gzhu.edu.cn}
\affiliation{
  School of Math and Information Science, Guangzhou University, Guangzhou 510006, China.}
\affiliation{
Guangdong Provincial Key Laboratory co-sponsored by province and city of Information Security Technology, Guangzhou University, Guangzhou 510006, China.}
\affiliation{
Guangzhou Center for Applied Mathematics, Guangzhou University, Guangzhou 510006, China.}
%\footnote{Author to whom correspondence should be addressed: pengjh@gzhu.edu.cn}

\author{Zhenhua Yuan}
\affiliation{
  School of Math and Information Science, Guangzhou University, Guangzhou 510006, China.}
\affiliation{
Guangdong Provincial Key Laboratory co-sponsored by province and city of Information Security Technology, Guangzhou University, Guangzhou 510006, China.}
\affiliation{
Guangzhou Center for Applied Mathematics, Guangzhou University, Guangzhou 510006, China.}
%\inst{1,2}\footnote{\email{pengjh@gzhu.edu.cn}},
%\shortauthor{J. H. Peng}

\date{\today}

\begin{abstract}
Random walks on directed networks provide a powerful framework for network search and exploration, yet most existing strategies follow edge directions, limiting access to upstream topology without explicit edge reversal.
{ To bridge this gap, we propose the reverse-graph stationary-probability-dependent random walk (RSRW) by leveraging the time reversal of the Markov chain associated with the classical random walk (CRW) on the reverse graph.}
We prove that RSRW preserves the stationary distribution, mean trapping times, and Kemeny's constant of CRW on the reverse graph, while strictly adhering to the original edge directions.
Numerical analysis \textbf{of} synthetic directed Erd\H{o}s--R\'{e}nyi networks and our newly proposed directed scale-free networks confirms the aforementioned theoretical findings, revealing that CRW preferentially targets high-in-degree nodes, whereas RSRW favors high-out-degree nodes.
Validation on the arXiv High Energy Physics citation network (Cit-HepTh) corroborates these findings: RSRW efficiently identifies review articles and interdisciplinary bridging papers.
Our results establish RSRW as an alternative strategy for directed network exploration in systems where counter-directional traversal is physically or operationally prohibited.
\end{abstract}

\pacs{05.40.Fb, 89.75.Hc, 05.60.Cd}
%\pacs{05.40.Fb}{Random walks and Levy flights}
%\pacs{05.60.Cd}{Classical transport}
%\pacs{89.75.Hc}network
%05.45.Df{Fractals}

\maketitle

%\begin{quotation}
\textbf{
Random walks serve as a paradigmatic model for transport and diffusion phenomena on complex networks, underpinning our understanding of processes as diverse as target search, reaction kinetics, network exploration, and more. While the theory is well established for undirected graphs, the inherently asymmetric nature of interactions in empirical systems—such as the World Wide Web, lymphatic vascular networks, and citation networks—necessitates a deeper understanding of directed dynamics. A persistent challenge in the field lies in the fact that existing strategies for random  walks on directed graphs predominantly operate in the \textit{downstream} direction, propagating along the natural orientation of directed edges. Consequently, accessing key \textit{upstream topological statistics} typically requires reversing all directed edges, which is difficult in many real-world applications. To bridge this gap, {we propose the reverse-graph stationary-probability-dependent random walk (RSRW). Constructed via the time reversal of the Markov chain characterizing CRW on the reverse graph, RSRW can access key upstream topological statistics—such as the stationary distribution, mean trapping times, and Kemeny's constant—without reversing any directed edges.}
We conduct numerical analyses on two synthetic networks and an arXiv citation network. The results confirm the aforementioned theoretical findings: CRW preferentially targets high-in-degree nodes, whereas RSRW favors high-out-degree nodes. Consequently, RSRW offers an alternative strategy for exploring directed systems, particularly those where counter-directional traversal is strictly prohibited due to physical or operational constraints.
}
%\end{quotation}

\section{Introduction}
\label{intro}
%反向拓扑探索的重要性，现有策略探索反向拓扑时都需要对图进行实际的反转，会因为成本太高或没有权限，导致不可能实施。

As a paradigmatic model of stochastic transport on discrete structures, random walks on graphs provide the theoretical bedrock for a broad class of  dynamical processes, spanning statistical mechanics, chemical kinetics, biological networks, and sociophysics \cite{Condamin-Benichou-Tejedor-2007-Nature, Masuda-2017-PR, Noh-Rieger-2004-PRL}. Key manifestations include  target search \cite{Benichou-PRL-2009, Sims-Southall-Humphries-Hays-2009-Nature}, reaction-diffusion processes on complex topologies \cite{Benichou-Chevalier-2010-NatChem}, community detection~\cite{Rosvall-Bergstrom-2008-PNAS, Rosvall-2009-map}, network optimization \cite{Peng-Chen-2022-TNSE, yuan2024fractal} and  exploration \cite{Page1999PageRank, Brin1998Anatomy, Prignano2012,  Kim2016}. On directed graphs\textemdash which can capture the inherent asymmetry of real-world interactions \cite{Stelling2002-Nature, Almaas2004-Nature, Price1965-science}\textemdash random walks can also serve as the generative mechanism for  graph representation learning~\cite{Perozzi2014DeepWalk} and recommender systems~\cite{Fouss2007RandomWalk,  Jamali-2009-trustwalker}.

The stochastic dynamics of random walks are governed not only by the underlying network topology~\cite{LO93,  Hwang-Lee-Kahng-2012-PRL, yuan2024first, yuan2024fractal} but also crucially by the transition rules defining the random walk strategy~\cite{Burda-2009-MERW, Tavani-2016-PRE, Lin-Zhang-2013-PRE, GaoPeng-2021-Jstat, Niu2018-MB, Peng-Zhang-2014-JCP}. Related  walking strategies include classical random walks (CRW)~\cite{ Hwang-Lee-Kahng-2012-PRL}, maximum entropy random walks~\cite{Burda-2009-MERW,  Peng-Zhang-2014-JCP},  edge-weight-based random walks~\cite{Zhang-2013-Weighted, GaoPeng-2021-Jstat},  resetting random walks~\cite{De-PRL-2020} \textbf{etc.}
The classical random walk, which hops to any neighboring node with uniform probability, offers a simple yet powerful probe of network structure~\cite{Rosvall-Bergstrom-2008-PNAS}. Investigations into CRW dynamics not only elucidate fundamental aspects of network connectivity~\cite{Johnson-2019-CSDA, LO93}, but also disclose the efficiency of information diffusion and transport across the system~\cite{Yuan2025-Chaos, Masuda-2017-PR, YuanPengGao2025-arXiv}.
Furthermore, the analytical framework established for CRW forms the foundational basis upon which more sophisticated random walk strategies are constructed and evaluated~\cite{Masuda-2017-PR, Burda-2009-MERW, Agliari-2012-NJP}. The maximum entropy random walk employs transition probabilities derived from the principal eigenvector of the adjacency matrix. This strategy maximizes the entropy rate of the path ensemble~\cite{gomez-2008-entropy, Burda-2009-MERW} and serves as a mechanism for enhancing transport efficiency across the network~\cite{Niu2018-MB, Peng-Zhang-2014-JCP}. Edge-weight-based random walks employ transition probabilities derived from network edge weights, offering a means to control and optimize transport efficiency~\cite{Zhang-2013-Weighted, Lin-Zhang-2013-PRE, GaoPeng-2021-Jstat}. Resetting random walks allow the walker to return to a designated location and restart the process, either probabilistically or according to a specific strategy. This mechanism enables the searcher to escape local optima, thereby significantly enhancing global search efficiency~\cite{De-PRL-2020, Wang-2021-CHAOS, Riascos-2020-PRE}. These random walk strategies each possess distinct dynamical characteristics, enriching the toolkit for probing complex networks.

{ Reverse-topology exploration has wide applications in source identification and backtracking \cite{Shen-2016-PhysRevE, Xie-Sekar-2005}. Examples include lymphoscintigraphy and sentinel lymph node mapping \cite{Kaushik2023, Martin2025, Breslin2019}, first-pass radionuclide angiography \cite{Friedman2006}, leak source localization and tracer-based inverse estimation in recirculating industrial cooling water systems \cite{GUO2026118867, COLOMBO2009212}, source tracking of illegal discharges in sewage and stormwater drainage networks \cite{HACHAD2022152060}, and the identification of initial triggering events in cascading failures \cite{crucitti2004-PRE, fekri2022identification}.
However, existing strategies for random walks on directed graphs, such as classical random walks, maximum entropy random walks, and edge-weight-based random walks, predominantly operate in the \textit{downstream} direction, propagating along the natural orientation of directed edges.
Consequently, accessing key \textit{upstream topological statistics} (such as the stationary distribution, mean trapping times, and Kemeny's constant\textbf{s}) typically requires reversing all directed edges.
However, in real engineering settings, many practical networks are subject to physical or operational constraints that strictly forbid any retrograde (against-the-arrow) movement.
Tracing or exploring can only advance along the original edge direction, never backwards \cite{Breslin2019, Kaushik2023, Friedman2006, COLOMBO2009212, HACHAD2022152060}\textbf{.}
This creates a compelling practical demand for mechanisms capable of achieving effective reverse-topological exploration without altering the original edge orientation.
To address this demand, we propose a new random walk strategy tailored for directed graphs, which is constructed via the temporal inversion of the Markov chain characterizing CRW on the reverse graph and can access the key upstream topological statistics without reversing any directed edges.}
The transition probability matrix of the proposed strategy is derived from the stationary distribution of the CRW on the reverse graph; accordingly, we term it \textit{Reverse-graph Stationary-probability-dependent Random Walk} (RSRW). We  analyze its stationary distribution and develop analytical methods for its mean first-passage time (MFPT) and mean trapping time (MTT). Furthermore, we perform a comparative analysis of RSRW and conventional CRW on two synthetic topologies and an arXiv citation network.

Our results demonstrate that although RSRW operates entirely on the original directed graph, its stationary distribution,  mean trapping times, and Kemeny's constant are identical to those of the CRW on the corresponding reverse graph. Thus, without altering the direction of edges in the original network, one can obtain  key topological and dynamic statistics  equivalent to those of the reverse graph's CRW. On directed networks exhibiting a significant disparity between node in-degrees and out-degrees, the stationary distributions and mean trapping times of CRW on the original and reverse graphs differ substantially. This new walk mechanism offers a fundamentally different approach and effect for exploring directed networks. Comparatively, CRW yields smaller MTTs at nodes with high in-degree, whereas RSRW achieves smaller MTTs at nodes with high out-degree.
Consequently, when applied to directed network exploration, CRW tends to more readily discover nodes with large in-degrees, while RSRW is more effective at locating nodes with large out-degrees. %These conclusions are validated through numerical analysis on both directed Erd\H{o}s--R\'{e}nyi  and directed scale-free networks.

The remainder of this paper is organized as follows. Sec.~\ref{sec:preliminaries} introduces the basic notation and theoretical tools used throughout the paper, including directed graphs, reverse graphs, finite-state Markov chains, and their first-passage quantities. Sec.~\ref{sec:random-walks-directed-graphs} presents the CRW and the  RSRW mechanism on directed graphs, together with comparisons of CRW on the reverse graph and RSRW on the original graph. Sec.~\ref{sec:random-walks-directed-er} investigates the two walking mechanisms on directed Erd\H{o}s--R\'{e}nyi networks, focusing on stationary distributions and the effect of in-degree (or out-degree) on MTTs. Secs. \ref{sec:DSF} and \ref{sec:cite_network} extend the analysis to a newly proposed directed scale-free network and an arXiv citation network. Finally, the main results are summarized in the concluding section. Additional derivations and supporting material are provided in the Appendices.

\section{Preliminaries}
\label{sec:preliminaries}

\subsection{Directed graphs and their reverse graphs}
\label{subsec:directed-reverse-graphs}
Let $G=(V,E)$ be a directed graph with $N=|V|$ nodes, where $V=\{v_1,v_2,\ldots,v_N\}$ is the node set of $G$, and $E\subseteq V\times V$ is the set of directed edges. There exists a directed edge in $G$ from node $i$ to node $j$ if and only if $(i,j)\in E$. The network structure is described by the adjacency matrix $A=\left(a_{ij}\right)_{N\times N}$, defined as
\begin{equation}
a_{ij}=
\begin{cases}
1, & \text{if } (i,j)\in E,\\
0, & \text{otherwise}.
\end{cases}
\end{equation}
Unlike the undirected case, the adjacency matrix $A$ of a directed graph is not necessarily symmetric, because the existence of a directed edge $(i,j)$ in $G$ does not imply the existence of its reverse edge $(j,i)$. %Unlike the undirected case, the adjacency matrix $A$ of a directed graph is not necessarily symmetric, because the existence of a directed edge $(i,j)\in E$ does not imply directed edge $(j,i)\in E$.

Accordingly, the out-degree $d_{\mathrm{out}}(i)$ and in-degree $d_{\mathrm{in}}(i)$ of an arbitrary node $i$ are defined, respectively, by
\begin{equation}
d_{\mathrm{out}}(i)=\sum_{j=1}^{N}a_{ij},\quad
d_{\mathrm{in}}(i)=\sum_{j=1}^{N}a_{ji}.
\end{equation}
Here, $d_{\mathrm{out}}(i)$ denotes the number of directed edges emanating from node $i$, while $d_{\mathrm{in}}(i)$ denotes the number of directed edges pointing to node $i$. Since this paper is primarily concerned with strongly connected directed graphs, every node $i$ satisfies $d_{\mathrm{out}}(i)\geq 1$ and $d_{\mathrm{in}}(i)\geq 1$.

For any directed graph $G=(V,E)$, graph $G^r=\left(V,E^r\right)$ is called the reverse graph of $G$ if and only if $E^r=\left\{(j,i)\mid (i,j)\in E,\ i,j\in V\right\}$. This means that the reverse graph is obtained by reversing the direction of every directed edge in the original graph. %Therefore, the reverse graph preserves the node set and the total number of edges of the original graph, but completely exchanges the direction of each directed edge.
Let $A^{RG}=\left(a_{ij}^{RG}\right)_{N\times N}$ denote the adjacency matrix of $G^r$. Then
\begin{equation}\label{eq:A-RG}
A^{RG}=A^T.
\end{equation}
For any node $i$ of the graph $G^r$, let $d_{out}^{RG}\left(i\right)$ and $d_{in}^{RG}\left(i\right)$ denote the out-degree and in-degree of node $i$ in the reverse graph $G^r$, respectively. We have
%The most direct structural relation between the original graph and its reverse graph is the interchange of in-degrees and out-degrees. Specifically, if the out-degree and in-degree of node $i$ in the reverse graph are denoted by $d_{\mathrm{out}}^{RG}\left(i\right)$ and $d_{\mathrm{in}}^{RG}\left(i\right)$, respectively, then
\begin{equation}\label{eq:d-RG}
d_{out}^{RG}\left(i\right)=d_{in}\left(i\right),\quad
d_{in}^{RG}\left(i\right)=d_{out}\left(i\right).
\end{equation}
Therefore, the reverse graph preserves the node set and the total number of edges of the original graph, but completely exchanges, for each node, the roles of receiving and sending connections. If $G$ is an undirected graph, $G=(V,E)$ and its own reverse graph $G^r=\left(V,E^r\right)$ are identical.
%Since the proposed mechanism explicitly uses information from the reverse graph, this notation will be used throughout the subsequent analysis.

\subsection{Definition and key properties for discrete-time and finite-state Markov chains}
\label{subsec:definition-key-metrics-markov-chains}
Let $I=\{i,j,k,\ldots\}$ be a finite state space, i.e., $N=|I|$ is a finite positive integer, and let $\{X_t\}_{t\geq 0}$ be a discrete-time Markov chain on $I$ with transition probabilities
\begin{equation}
\mathbb{P}(X_{t+1}=j \mid X_t=i) \triangleq p_{ij}, \quad i,j\in I.
\end{equation}
The matrix $P=\left(p_{ij}\right)_{N\times N}$ is called the transition matrix of the Markov chain.

A probability vector $\vec{\pi}=\left(\pi_1,\pi_2,\ldots,\pi_N\right)$ is called a stationary distribution of the chain if and only if
\begin{equation}\label{stationary-equ}
\vec{\pi}P=\vec{\pi}, \qquad \sum_{i=1}^{N}\pi_i=1, \qquad \pi_i>0.
\end{equation}
If $\{X_t\}_{t\geq 0}$ is a finite irreducible Markov chain, the stationary distribution $\vec{\pi}$ exists and is unique.

For any state $j\in I$, the first passage time (FPT) from state $i$ to state $j$ $(i\neq j)$, denoted as $\mathrm{FPT}_{ij}$, is defined by
\begin{equation}
\mathrm{FPT}_{ij}=\min\{t|X_0=i, X_t=j, t>0\}.
\end{equation}
Then, the mean first-passage time (MFPT) from state $i$ to state $j$ ($i \neq j$), denoted by $T_{ij}$, is defined as
\begin{equation}\label{eq:8}
T_{ij} = \mathbb{E}\left[\mathrm{FPT}_{ij}\right],
\end{equation}
where $\mathbb{E}\left[\mathrm{FPT}_{ij}\right]$ denotes the expectation of the random variable $\mathrm{FPT}_{ij}$.
When $i = j$, $T_{ij}$ is referred to as the mean first return time, denoted by $\mathrm{MRT}_j$.
When the matrix $T = \left(T_{ij}\right)_{N \times N}$ is expanded, its diagonal entries are conventionally set to $T_{jj} = \mathrm{MRT}_j$.

%Then, the mean first-passage time (MFPT) from state $i$ to state $j$ $(i\neq j)$, denoted by $T_{ij}$, is defined as
%\begin{equation}\label{eq:8}
%T_{ij}=\mathbb{E}\left[\mathrm{FPT}_{ij}\right],
%\end{equation}
%where $\mathbb{E}\left[\mathrm{FPT}_{ij}\right]$ denotes the expectation of the random variable $\mathrm{FPT}_{ij}$. In the case where $i=j$, $T_{ij}$ turns to be the mean first return time, denoted by $\mathrm{MRT}_j$. When the matrix $T=\left(T_{ij}\right)_{N\times N}$ is written in full, its diagonal entries may be conventionally set as $T_{jj}=\mathrm{MRT}_j$.

For each target state $j$, the mean trapping time (MTT), defined by
\begin{equation}\label{mtt}
\mathrm{MTT}_j=\sum_{i=1}^{N}{\pi_iT_{ij}}
\end{equation}
is the stationary-weighted average of $T_{ij}$ over all source states $i$.

{
The Kemeny's constant of the Markov chain is defined by
\begin{equation}\label{Kemeny}
K=\sum_{j=1}^{N}\sum_{i\ne j}{\pi_i\pi_jT_{ij}},
\end{equation}
and can serve as a measure for transport (search) efficiency of the system.}

Random walks on a directed graph correspond to discrete-time finite-state Markov chains. If the directed graph is finite and strongly connected, every site of the graph can be reached from any other site, and the induced chain associated with the random walk process is therefore finite and irreducible. Under this condition, the chain admits a unique stationary distribution, which is strictly positive on every state. Hence, the stationary distribution provides a fundamental description of the long-term statistical structure of the random walk. Similarly, the mean first passage time, the mean trapping time, and Kemeny's constant for random walks on directed graphs can also be defined by using Eqs.~\eqref{eq:8}--~\eqref{Kemeny}.

\section{Random Walks on Directed Graphs}
\label{sec:random-walks-directed-graphs}

\subsection{Classical random walk strategy}
\label{subsec:classical-random-walk}
Let $G=\left(V,E\right)$ be a finite strongly connected directed graph, and let $A=\left(a_{ij}\right)_{N\times N}$ denote its adjacency matrix. The classical random walk (CRW) on graph $G$ can be depicted by the following one-step transition probability,
\begin{equation}\label{CRW-p}
p_{ij}^{\mathrm{CRW}}=
\begin{cases}
\dfrac{a_{ij}}{d_{\mathrm{out}}(i)}, & d_{\mathrm{out}}(i)\ne 0,\\[6pt]
0, & d_{\mathrm{out}}(i)=0,
\end{cases}
\end{equation}
where $d_{\mathrm{out}}\left(i\right)$ and $d_{\mathrm{in}}\left(i\right)$ represent the out-degree and in-degree of node $i$ in graph $G$, respectively. That is to say, at each step, the walker, located at node $i$, selects uniformly from all out-neighbors of $i$, and the walker then moves along the corresponding directed edge.

Let $P^{\mathrm{CRW}}=\left(p_{ij}^{\mathrm{CRW}}\right)_{N\times N}$. Then Eq.~\eqref{CRW-p} can be rewritten as
\begin{equation}
P^{\mathrm{CRW}}=\left(D_{\mathrm{out}}\right)^{-1}A,
\end{equation}
where $D_{\mathrm{out}}$ is the diagonal matrix of node out-degrees, namely $D_{out}= diag(d_{out}(1),d_{out}(2),...,d_{out}(N))$.

This definition shows that CRW depends only on the local forward adjacency structure of the original graph. %: the next-step choice is determined entirely by the distribution of outgoing edges from the current node. No additional weight reassignment is involved, and no information from the reverse graph is used.
%Therefore, this mechanism describes the natural diffusion of a walker on the original directed network and provides a direct reference process for the new random-walk mechanism to be introduced later.

Since all networks considered here are finite strongly connected directed graphs, CRW defined by $P^{CRW}$ forms a finite irreducible Markov chain. It therefore possesses a unique stationary distribution~\cite{Masuda-2017-PR}
\begin{equation}
\vec{\pi}^{CRW}=\left(\pi_1^{CRW},\pi_2^{CRW},...,\pi_N^{CRW}\right),
\end{equation}
satisfying
\begin{equation}
\vec{\pi}^{CRW}P^{CRW}=\vec{\pi}^{CRW},\qquad
\sum_{i=1}^{N}\pi_i^{CRW}=1,\qquad
\pi_i^{CRW}>0.
\end{equation}

Let $G^r=\left(V,E^r\right)$ denote the reverse graph, as defined in Sec.~\ref{subsec:directed-reverse-graphs}, of graph $G=\left(V,E\right)$.
The transition probability of the CRW on $G^r$ from node $i$ to node $j$ is given by %By using Eqs.~\eqref{eq:A-RG}--\eqref{eq:d-RG},
\begin{equation}\label{eq:RG-p-AD}
p_{ij}^{RG}=\frac{A_{ij}^{RG}}{d_{out}^{RG}\left(i\right)}=\frac{A_{ji}}{d_{in}\left(i\right)}
\end{equation}
where $A_{ji}$ denotes the $(j,i)$-th entry of the adjacency matrix $A$ of the original graph $G$, and $d_{\mathrm{in}}\left(i\right)$ denotes the in-degree of node $i$ in $G$.

Let $P^{RG}={(p_{ij}^{RG})}_{N\times N}$. Then Eq.\eqref{eq:RG-p-AD} can be rewritten as
\begin{equation}
P^{RG}=D_{in}^{-1}A^T=\left(AD_{in}^{-1}\right)^T,
\label{P_RG}
\end{equation}
where $D_{in}=diag\left(d_{in}\left(1\right),d_{in}\left(2\right),\ldots,d_{in}\left(N\right)\right)$ is the diagonal matrix formed by the in-degrees of the nodes in $G$.

If $G$ is a finite strongly connected directed graph, the reverse graph $G^r$ is also a finite strongly connected directed graph, and CRW on $G^r$ forms a finite irreducible Markov chain. It therefore possesses a unique stationary distribution, denoted by $\vec{\phi}=\left(\phi_1,\ldots,\phi_N\right)$, which satisfies
\begin{equation}\label{eq:RG-stationary-distribution}
\vec{\phi}P^{RG}=\vec{\phi},\qquad
\sum_{i=1}^{N}\phi_i=1,\qquad
\phi_i>0.
\end{equation}
In general, the reverse graph $G^r$ has a different structure from that of the original graph $G$, and $\vec{\phi}\neq \vec{\pi}^{CRW}$.

\subsection{Reverse-graph Stationary-probability-dependent Random Walk on directed graphs}
%{A novel random walk strategy on directed graphs}
\label{subsec:RSRW}
%We now introduce a novel random-walk strategy on  directed graph. %The the transition probability
%It is constructed so that its stationary distribution coincides with that of the classical random walk on the reverse graph. We therefore refer to it as the Reverse-graph Stationary-probability-dependent Random Walk (RSRW).

%Let $G=\left(V,E\right)$ be a finite strongly connected directed graph with adjacency matrix $A=\left(A_{ij}\right)_{N\times N}$, $G^r=\left(V,E^r\right)$ denote its reverse graph defined in Sec.~\ref{subsec:directed-reverse-graphs}, and $\vec{\phi}=\left(\phi_1,\ldots,\phi_N\right)$ be the stationary distribution for CRW on reverse graph $G^r$. By leveraging the time reversal of the Markov chain associated with CRW on the reverse graph $G^r$, we obtain a new random walk on directed graph $G$, which can be depicted by the transition matrix $P^{RSRW}=\left(p_{ij}^{RSRW}\right)_{N\times N}$, where
%\begin{equation}\label{RSRW-p}
%  p_{ij}^{RSRW}=\frac{\phi_j}{\phi_i}\frac{A_{ij}}{d_{in}\left(j\right)},
%\end{equation}
%$A_{ij}$ denotes the $(i,j)$-th entry of the adjacency matrix $A$ of the original graph $G$, and $d_{\mathrm{in}}\left(j\right)$ denotes the in-degree of node $j \in V$ in $G$. Because the the transition probabilities for this new random walk are derived from the stationary distribution for  CRW on its reverse graph. We term it the Reverse-graph Stationary-probability-dependent Random Walk (RSRW).
{
Let $G=(V,E)$ be a finite strongly connected directed graph with adjacency matrix
$A=(A_{ij})_{N\times N}$, and let $G^r=(V,E^r)$ denote its reverse graph (Sec.~\ref{subsec:directed-reverse-graphs}).
Let $\vec{\phi}=(\phi_1,\ldots,\phi_N)$ be the stationary distribution of CRW on $G^r$.
Leveraging the time-reversed Markov chain of CRW on $G^r$, we define a new walk on $G$
with transition matrix $P^{\text{RSRW}}=(p_{ij}^{\text{RSRW}})_{N\times N}$ given by
\begin{equation}\label{RSRW-p}
  p_{ij}^{\text{RSRW}}=\frac{\phi_j}{\phi_i}\,\frac{A_{ij}}{d_{\mathrm{in}}(j)},
\end{equation}
where $A_{ij}$ is the $(i,j)$ entry of $A$ and $d_{\mathrm{in}}(j)$ is the in-degree of $j\in V$ in $G$.
As the transition probabilities are sourced from the stationary distribution of CRW on $G^r$,
we refer to this walk as the \textit{reverse-graph stationary-probability-dependent random walk} (RSRW).
}

One can also rewrite Eq.~\eqref{RSRW-p} as
\begin{equation}\label{eq:24}
P^{RSRW}={\widetilde{\mathrm{\Phi}}}^{-1}A\operatorname{D}_{in}^{-1}\widetilde{\mathrm{\Phi}},
\end{equation}
where $\widetilde{\mathrm{\Phi}}=\operatorname{diag}\left(\phi_1,\ldots,\phi_N\right)$ is the diagonal matrix formed by the stationary distribution of CRW on the reverse graph $G^r$,  and $D_{in}=\operatorname{diag}\left(d_{in}\left(1\right),d_{in}\left(2\right),\ldots,d_{in}\left(N\right)\right)$ is the diagonal matrix composed of the in-degrees of the nodes in $G$.

It is easy to know that $p_{ij}^{RSRW} \geq 0$ for all $i,j\in V$, and for any $i\in V$,
\begin{equation}
\phi_i=\left(\vec{\phi}P^{RG}\right)_i=\sum_{j=1}^{N}{\phi_jP_{ji}^{RG}}=\sum_{j=1}^{N}{\phi_j\frac{A_{ij}}{d_{in}\left(j\right)}}\ .
\end{equation}
Thus
\begin{equation}
\sum_{j=1}^{N}p_{ij}^{RSRW}=\sum_{j=1}^{N}{\frac{\phi_j}{\phi_i}\frac{A_{ij}}{d_{in}\left(j\right)}}=1,
\end{equation}
which implies the transition probabilities $p_{ij}^{RSRW}$ are well defined.

%One can also rewrite Eq.~\eqref{RSRW-p} as
%\begin{equation}\label{eq:24}
%P^{RSRW}={\widetilde{\mathrm{\Phi}}}^{-1}AD_{in}^{-1}\widetilde{\mathrm{\Phi}},
%\end{equation}
%where $\widetilde{\mathrm{\Phi}}=diag\left(\phi_1,\ldots,\phi_n\right)$ is the diagonal matrix formed by the stationary distribution of CRW on the reverse graph $G^r$,  and $D_{in}=diag\left(d_{in}\left(1\right),d_{in}\left(2\right),\ldots,d_{in}\left(N\right)\right)$ is the diagonal matrix composed of the in-degrees of the nodes in $G$.

%It is easy to know that $p_{ij}^{RSRW} \geq 0$ for all $i,j\in V$, and for any $i\in V$,
%\begin{equation}
%\phi_i=\left(\vec{\phi}P^{RG}\right)_i=\sum_{j=1}^{N}{\phi_jP_{ji}^{RG}}=\sum_{j=1}^{N}{\phi_j\frac{A_{ij}}%{d_{in}\left(j\right)}}\ .
%\end{equation}
%Thus
%\begin{equation}
%\sum_{j=1}^{N}p_{ij}^{RSRW}=\sum_{j=1}^{N}{\frac{\phi_j}{\phi_i}\frac{A_{ij}}{d_{in}\left(j\right)}}=1,
%\end{equation}
%which implies the transition probabilities $p_{ij}^{RSRW}$ are well defined.

Substituting Eq.~\eqref{P_RG} into Eq.~\eqref{eq:24}, we obtain
\begin{equation}
P^{\mathrm{RSRW}}
=
\widetilde{\mathrm{\Phi}}^{-1}
\bigl(P^{\mathrm{RG}}\bigr)^{\!T}
\widetilde{\mathrm{\Phi}},
\label{eq:PRSRW}
\end{equation}
{ which implies that the Markov chain associated with RSRW on graph $G$
is the time reversal of the Markov chain associated with CRW on the reverse graph $G^r$.
}

%For an undirected graph $G$, namely, when the adjacency matrix is symmetric (i.e.,$\ A=A^T$), one readily verifies that RSRW defined in Eq.~\eqref{RSRW-p} reduces to CRW.
%On general directed graphs with non-symmetric adjacency matrices, RSRW acts on the original graph and CRW on its reverse  counterpart. Their induced trajectories differ completely, marking them as distinct random walk paradigms. %Since their transition probability matrices have close relation, we will show that they have \ref{subsec:Comparison_RSRW_CRW}

For an undirected graph $G$, specifically when the adjacency matrix is symmetric (i.e., $A=A^T$), it can be readily verified that the RSRW defined in Eq.~\eqref{RSRW-p} reduces to CRW.
In contrast, on general directed graphs with non-symmetric adjacency matrices, RSRW acts on the original graph, while CRW acts on its reverse counterpart. Their induced trajectories differ completely, marking them as distinct
random-walk strategies.

\subsection{Methods for evaluating the MFPT and MTT for random walks on directed graphs}
\label{subsec:method_MFPT_MTT}
In this section, we present a unified approach for evaluating the mean first-passage time (MFPT) and mean trapping time (MTT) for the random walks considered in this paper. The derivation is carried out at the level of a finite irreducible Markov chain and therefore does not depend on the specific form of the transition rule. Consequently, the results apply equally to CRW on the original directed graph and to RSRW introduced above. This framework provides the main computational basis for the subsequent analysis and enables a consistent comparison between the two walk mechanisms.

%For a finite irreducible Markov chain with transition matrix $P$ and stationary distribution $\vec{\pi}=\left(\pi_1,\ldots,\pi_N\right)$, let $\vec{1}$ denote a column vector with all entries  equal to $1$ and define
%\begin{equation}
%\Pi=\vec{1}\vec{\pi}.
%\end{equation}
%The fundamental matrix associated with $P$ is introduced as~\cite{Kemeny-1983-fundamental-matrix}
%\begin{equation}
%Z=\left(I-P+\Pi\right)^{-1}.
%\end{equation}
%Here, matrix $I$ is the identity matrix. For a finite irreducible Markov chain, the matrix $I-P+\Pi$ is nonsingular, and hence $Z$ is well defined. This matrix provides a convenient analytical tool for characterizing first-passage-related quantities in a unified manner.

Let $P$ be the transition matrix for random walks on a finite strongly connected directed graph $G$ with node set  $V=\{v_1,v_2,\ldots,v_N\}$,
 $\vec{\pi}=\left(\pi_1,\pi_2,\ldots,\pi_N\right)$ its stationary distribution, $\vec{1}$ a column vector with all entries equal to $1$, $\Pi=\vec{1}\vec{\pi}$, and $I$ the identity matrix. The fundamental matrix associated with $P$ can be written  as~\cite{Kemeny-1983-fundamental-matrix}
\begin{equation}
Z=\left(I-P+\Pi\right)^{-1}.
\end{equation}
%where detailed information for the fundamental matrix is collected in Appendix \ref{subsec:fundamental-matrix}.

% Denote by $Z=\left(z_{ij}\right)_{N\times N}$ the corresponding fundamental matrix defined in ec.~\ref{subsec:fundamental-matrix}.
Then, for any two nodes $i$ and $j$ with $i\neq j$, the MFPT from $i$ to $j$ can be expressed as~\cite{Kemeny-1983-fundamental-matrix}
\begin{equation}
T_{ij}=\frac{z_{jj}-z_{ij}}{\pi_j}.
\end{equation}
For $i=j$, the diagonal entries of the MFPT matrix $T=\left(T_{ij}\right)_{N\times N}$ are taken to be the mean return times (MRTs)~\cite{Kemeny-1983-fundamental-matrix}, namely,
\begin{equation}
T_{jj}=\frac{1}{\pi_j}.\
\end{equation}
Accordingly, the MFPT matrix $T$ can be written in the unified form
\begin{equation}\label{eq:30}
T_{ij}=\frac{z_{jj}-z_{ij}+\delta_{ij}}{\pi_j}\ ,\ \ \forall i,j\in V\ ,
\end{equation}
where $\delta_{ij}$ is the Kronecker delta. %This expression will be used throughout the subsequent analysis.

Let $E=\vec{1}{\vec{1}}^T$ denote the matrix whose entries are all equal to 1, and let $Z_{dg}=diag\left(z_{11},z_{22},\ldots,z_{NN}\right)$. Then the above expression can be written in matrix form as
\begin{equation}\label{eq:31}
T=\left(I-Z+EZ_{dg}\right)D^{-1},
\end{equation}
where $D$ is the diagonal matrix formed by the components of the stationary distribution $\vec{\pi}$, namely, $D=diag(\pi_1,\pi_2,\ldots,\pi_N)$.

Eq.~\eqref{eq:31} shows that, once the transition matrix $P$ is given, the entire MFPT matrix can be computed uniformly from the stationary distribution and the fundamental matrix. Hence, this method is applicable to both CRW ($P=P^{CRW}$) and  RSRW ($P=P^{RSRW}$).

We next derive the MTT from the MFPT. For convenience, we recall the definition in Eq.~\eqref{mtt}. For a fixed target node $j\in V$,
\[
\mathrm{MTT}_j=\sum_{i=1}^{N}\pi_i T_{ij},
\]
where $T_{ij}$ denotes the MFPT from node $i$ to node $j$. By using Eq.~\eqref{eq:30}, one has
\begin{equation}
\mathrm{MT}\mathrm{T}_\mathrm{j}=1+\sum_{i\neq j}{\pi_i\frac{z_{jj}-z_{ij}}{\pi_j}}.
\end{equation}
Since %From the fundamental-matrix identity in Eq.~(15)
\begin{equation}
\sum_{i=1}^{N}{\pi_iz_{ij}}=\pi_j,
\end{equation}
one gets
\begin{equation}
\sum_{i\neq j}{\pi_iz_{ij}}=\pi_j-\pi_jz_{jj}.
\end{equation}
Therefore, the MTT of node $j$ on a finite irreducible directed graph can be written as
\begin{equation}\label{eq:MTT-fm}
\mathrm{MT}\mathrm{T}_\mathrm{j}=\frac{z_{jj}}{\pi_j}.
\end{equation}
Eq.~\eqref{eq:MTT-fm} provides a compact expression for the MTT of node~$j$
for random walks on a finite, directed graph whose induced chain is irreducible. It shows that $\mathrm{MT}\mathrm{T}_j$ is jointly determined by the stationary weight $\pi_j$ of the target node and the diagonal entry $z_{jj}$ of the fundamental matrix. Since this expression depends only on the stationary distribution and the fundamental matrix associated with the transition matrix $P$, it remains valid for both CRW and RSRW.

Accordingly, all MFPT and MTT quantities considered in the following numerical analysis can be computed in a unified way directly from the corresponding transition matrix.

\subsection{Comparison of RSRW on the original graph and CRW on the reverse graph}
\label{subsec:Comparison_RSRW_CRW}

Let $G=(V,E)$ be a finite, strongly connected directed graph with adjacency matrix $A=(A_{ij})_{N\times N}$,
and let $G^r=(V,E^r)$ denote its reverse graph as defined in Sec.~\ref{subsec:directed-reverse-graphs}.
We now show, by employing the theory of time-reversed Markov chains,
that the stationary distribution, the mean trapping time for an arbitrary node,
and Kemeny's constant for RSRW on $G$ are identical to those for CRW on $G^r$,
although the mean first-passage time between arbitrary pairs of nodes generally differs for the two walks.

First, we consider the stationary distribution.
Let $\vec{\phi}=\left(\phi_1,\ldots,\phi_N\right)$ be the stationary distribution for CRW on the reverse graph $G^r$.
It is easy to verify from the definition of \(P^{\mathrm{RSRW}}\) that
\begin{equation}
\vec{\phi}P^{RSRW}=\vec{\phi}\ .
\end{equation}
Thus, $\vec{\phi}$ is also a stationary distribution for the RSRW on graph $G$. %Since the original graph $G$ is finite and strongly connected, the transition matrix $P^{RSRW}$ is irreducible. Therefore, the induced Markov chain is finite and irreducible. Then $\vec{\phi}$ is the unique stationary distribution for RSRW on directed graph $G$.

%For the reverse graph $G^r$, the transition matrix of CRW is
%\begin{equation}
%P^{RG}=D_{\mathrm{in}}^{-1}A^T
%       =\left(AD_{\mathrm{in}}^{-1}\right)^T.
%\label{eq:PRG}
%\end{equation}

Then we will show that, for an arbitrary node $i \in V$,
\begin{equation}
\mathrm{MTT}_i^{\mathrm{RSRW}}
=
\mathrm{MTT}_i^{\mathrm{RG}},
\label{Equv_MTT}
\end{equation}
where $\mathrm{MTT}_i^{\mathrm{RSRW}}$ and $\mathrm{MTT}_i^{\mathrm{RG}}$ denote the MTT of node $i$ for RSRW on $G$ and CRW on $G^r$, respectively.

Let $Z^{\mathrm{RSRW}}$ and $Z^{\mathrm{RG}}$ denote the fundamental matrices
corresponding to RSRW on $G$ and CRW on  $G^{r}$, respectively.
According to the relation between $\mathrm{MTT}_i$ and $Z_{ii}$ given in Eq.~(\ref{eq:MTT-fm}),
it suffices to show that for an arbitrary node $i \in V$,
\begin{equation}
Z_{ii}^{\mathrm{RSRW}}
=
Z_{ii}^{\mathrm{RG}},
\label{eq:Z_diag_equal}
\end{equation}
where the left-hand side is the $i$th diagonal entry of $Z^{\mathrm{RSRW}}$ (for RSRW on $G$)
and the right-hand side is the $i$th diagonal entry of $Z^{\mathrm{RG}}$ (for CRW on $G^{r}$), respectively.

In the following, we will show that Eq.~(\ref{eq:Z_diag_equal}) holds for an arbitrary node $i\in V$.

Let $\Pi=\vec{\mathbf{1}}\vec{\phi}$, where $\widetilde{\Phi}=\mathrm{diag}(\phi_1,\ldots,\phi_N)$. We get
\begin{equation}
\widetilde{\Phi}^{-1}\Pi^{T}\widetilde{\Phi}
=
\widetilde{\Phi}^{-1}
\vec{\phi}^{T}
\vec{\mathbf{1}}^{T}
\widetilde{\Phi}
=
\vec{\mathbf{1}}\vec{\phi}
=
\Pi ,
\end{equation}
and
\begin{equation}
I-P^{\mathrm{RSRW}}+\Pi
=
I-\widetilde{\Phi}^{-1}
\left(P^{\mathrm{RG}}\right)^{T}
\widetilde{\Phi}
+\Pi .
\end{equation}
Therefore,
\begin{align}
Z^{\mathrm{RSRW}}
&=
\left[
I-
\widetilde{\Phi}^{-1}
\left(P^{\mathrm{RG}}\right)^T
\widetilde{\Phi}
+\Pi
\right]^{-1} \nonumber\\
&=
\left[
\widetilde{\Phi}^{-1}
\left(
I-\left(P^{\mathrm{RG}}\right)^T+\Pi^T
\right)
\widetilde{\Phi}
\right]^{-1} \nonumber\\
&=
\widetilde{\Phi}^{-1}
\left[
I-\left(P^{\mathrm{RG}}\right)^T+\Pi^T
\right]^{-1}
\widetilde{\Phi} \nonumber\\
&=
\widetilde{\Phi}^{-1}
\left[
\left(I-P^{\mathrm{RG}}+\Pi\right)^{-1}
\right]^T
\widetilde{\Phi} \nonumber\\
&=
\widetilde{\Phi}^{-1}
\left(Z^{\mathrm{RG}}\right)^T
\widetilde{\Phi}.
\label{eq:Z_RSRW_Z_RG_relation}
\end{align}

Taking the diagonal entries of Eq.~\eqref{eq:Z_RSRW_Z_RG_relation}, we obtain Eq.~(\ref{eq:Z_diag_equal}).
%\begin{equation}
%Z_{ii}^{\mathrm{RSRW}}
%=
%\frac{1}{\phi_i} Z_{ii}^{\mathrm{RG}}\phi_i
%=
%Z_{ii}^{\mathrm{RG}},
%\label{eq:Z_diag_equal}
%\end{equation}
%for an arbitrary node $i \in V$.
Hence, $Z^{\mathrm{RSRW}}$ and $Z^{\mathrm{RG}}$ have identical diagonal elements, although their off-diagonal elements are generally different.

%Therefore, Eqs.~(\ref{eq:Z_diag_equal}) and (\ref{Equv_MTT}) are also hold. For an arbitrary node $i \in V$, The MTT of node $i$ for RSRW on $G$ is identical to that of CRW on the reverse graph $G^r$. Since Kemeny's constant is the weighted average for $\mathrm{MTT}_i$ over all the node $i$ of the network. 理所当然的是，the Kemeny's constant of the RSRW on $G$ is also identical to that of the CRW on the corresponding  reverse graph $G^r$. \eqref{eq:Z_diag_equal} and

Consequently, Eq.~ \eqref{Equv_MTT} holds.
For any node $i\in V$, the MTT for RSRW on $G$ coincides with that for CRW on $G^r$.
Since Kemeny's constant is the weighted average of $\mathrm{MTT}_i$ over all nodes of the graph,
the Kemeny constant of RSRW on $G$ equals that of CRW on the reverse graph $G^r$.

{ Although RSRW operates entirely on the original directed graph $G$, it yields stationary distributions, mean trapping times ($\mathrm{MTT}$), and Kemeny's constant identical to those of CRW on the reverse graph $G^r$. Since $\mathrm{MTT}$ and Kemeny's constant serve as canonical metrics for transport and search efficiency in networked systems, these results demonstrate that comparable transport efficiency can be attained without modifying the intrinsic edge orientations of $G$. This property is particularly consequential for direction-constrained networks---including lymphatic vascular networks, underground sewer networks and monorail metro lines with unidirectional track layouts---where counter-directional traversal is physically or operationally prohibited. Consequently, the RSRW establishes a new operational strategy for network exploration in such systems.

For most directed graphs, nodal stationary probabilities under CRW correlate
strongly and positively with in-degree \cite{Cooper-2012-JCTSB, Norris_1997, Caputo2020};
nodes with larger in-degrees thus exhibit higher stationary probabilities and,
correspondingly, shorter mean trapping times, making CRW biased toward
high-in-degree nodes.
Since a node's in-degree in $G^r$ equals its out-degree in $G$, and since
RSRW on $G$ shares the stationary distribution of CRW on $G^r$,
RSRW stationary probabilities are instead governed by out-degree in $G$—
nodes with larger out-degrees tend to have larger stationary probabilities
and shorter MTTs under RSRW, making RSRW more effective at surfacing
high-out-degree nodes.
For undirected graphs, RSRW reduces to CRW;
for directed graphs with substantial in-out imbalance, however,
RSRW transition behavior diverges markedly from CRW's.
 }

\section{Random walks on directed Erd\H{o}s--R\'{e}nyi networks}
\label{sec:random-walks-directed-er}

In this section, we use directed Erd\H{o}s--R\'{e}nyi networks as a case study.
We first validate numerically that the stationary distribution, the mean trapping time (MTT) and Kemeny's constant for RSRW on the original directed graph~$G$ coincide with those for CRW on the reverse graph~$G^{r}$ across various network configurations. Subsequently, we investigate how the MTT depends on the in-degree and out-degree of nodes for both CRW and RSRW on~$G$.

\subsection{Network construction}
\label{subsec:der-network-construction}

Directed Erd\H{o}s--R\'{e}nyi random graphs $G(N,p)$ are constructed from $N$ labeled nodes: for any pair $i\neq j$, a directed edge $i\rightarrow j$ is added independently with probability $p$, with self-loops excluded ($a_{ii}=0$)~\cite{Tishby-2017-JOSMTAE}. The network is encoded by the adjacency matrix $A=(a_{ij})_{N\times N}$, where $a_{ij}=1$ if $i\rightarrow j$ exists and $0$ otherwise. As $N$ grows, the probabilities of isolated nodes ($e^{-2p(N-1)}$) or nodes with only in-/out-going edges ($e^{-p(N-1)}$ each) decay exponentially. For $p$ exceeding the connectivity threshold, such anomalous nodes are negligible, yielding valid samples for random-walk analysis.

Crucially, when $p\gg \ln{N}/N$, the probability that the graph is strongly connected approaches unity~\cite{Frieze-2016-IRG}. Without loss of generality, we therefore assume that all directed Erd\H{o}s--R\'{e}nyi networks considered herein are finite and strongly connected. This ensures that the induced Markov chains for both CRW and RSRW are finite and irreducible, allowing for the rigorous evaluation of the stationary distribution, mean first-passage times, and mean trapping times for an arbitrary node $i$. %To guarantee strong connectivity of analyzed networks, we retain strongly connected realizations generated by the aforementioned method for each selected $(N,p)$ and discard non-strongly-connected ones.

\subsection{Equivalence of stationary and transient properties: RSRW on $G$ vs. CRW on $G^r$}
\label{subsec:DER_equivalence_validation}
{
In this subsection, we numerically validate the theoretical equivalence established in Sec.~\ref{subsec:Comparison_RSRW_CRW}: for any valid network configuration, the stationary distribution, mean trapping time (MTT) for an arbitrary node $i$, and Kemeny's constant of RSRW on the original graph $G$ coincide exactly with the corresponding metrics of CRW on the reverse graph $G^r$, while differing significantly from those of CRW on $G$.

We generate directed scale-free graphs $G$ under various
$(N,p)$ combinations
via the construction in Sec.~\ref{subsec:der-network-construction},
and form the corresponding reverse graph $G^r$ for each $G$.
%Fig.~\ref{fig:der-graph-compare-g-gr} illustrates a representative instance with $N=20$ and $p=0.15$.
%\begin{figure}[!b]
%    \centering
%    \includegraphics[width=\columnwidth]{DER-graph_compare_G_Gr.eps}
%    \caption{Structural comparison of a directed Erd\H{o}s--R\'{e}nyi network $G(N,p)$ (here referred to as $G$) with $N=20$ and $p=0.15$, and its reverse graph $G^r$.}  %The left and right panels show $G$ and $G^r$, respectively. Both graphs are drawn with the same node layout;  Red dots denote vertices and arrows indicate edge directions.
%    \label{fig:der-graph-compare-g-gr}
%\end{figure}

We construct the transition matrices for CRW on $G$, CRW on $G^r$, and RSRW on $G$
using Eqs.~\eqref{CRW-p}, \eqref{eq:RG-p-AD}, and \eqref{RSRW-p}, respectively.
Applying the method in Sec.~\ref{subsec:method_MFPT_MTT},
we compute for each walk the stationary distribution, the mean trapping time ($\mathrm{MTT}$) for an arbitrary node,
and Kemeny's constant.

Numerical results confirm that, across all tested network configurations, the stationary distribution, $\mathrm{MTT}$, and Kemeny's constant of RSRW on $G$ differ from those of CRW on $G$, but coincide exactly with the corresponding metrics of CRW on $G^r$. Detailed empirical evidence supporting the equivalence of the stationary distributions is provided in Appendix~\ref{appendix:der-stationary-distribution}.
Beyond the exact equivalence of global metrics established above, we further observe that among nodes with out-degree exceeding in-degree ($d_i^{out}>d_i^{in}$), the vast majority exhibit higher stationary probabilities under RSRW than under CRW. This indicates that RSRW preferentially assigns greater stationary weight to nodes with $d_i^{out}>d_i^{in}$. More broadly, this disparity in stationary allocation reflects a fundamental divergence in how the two walks redistribute stationary probability mass across the network.

%Appendix~\ref{appendix:DER-stationary-distribution} presents a representative directed Erd\H{o}s--R\'{e}nyi network ($N=20$, $p=0.15$) and the corresponding stationary distributions for CRW on $G$, CRW on $G^r$, and RSRW on $G$.

%\begin{figure}[!b]
%    \centering
%    \includegraphics[width=0.9\columnwidth]{DER-stationary_distribution_compare.eps}
 %   \caption{Scatter plot comparison for the three stationary distributions of the illustrative networks shown in Fig.~\ref{fig:der-graph-compare-g-gr}.
%     Blue squares denote  stationary probability for CRW on $G$, green dots denote stationary probabilities for CRW on $G^r$, red circles denote stationary probabilities for RSRW on $G$.
    %Stationary probabilities of the three random-walk strategies at each node. The calculations are performed on the directed graph $G$ shown in the left panel of Fig.~\ref{fig:der-graph-compare-g-gr}. Blue squares denote the classical random walk on the original graph, red circles denote the new random-walk strategy on the original graph, and green dots denote the classical random walk on the reverse graph.
%    \label{fig:der-stationary-distribution-compare}
%\end{figure}

}

\subsection{Degree dependence of mean trapping times}
\label{subsec:ER_degree_dependence}
In this subsection, we investigate how the MTT depends on the in-degree and out-degree of nodes for both CRW and RSRW on directed Erd\H{o}s--R\'{e}nyi networks.

{ The method we use to evaluate the MTT is shown in Subsection \ref{subsec:method_MFPT_MTT}.
To mitigate random fluctuations arising from individual network instances, we generate 100 valid realizations of $G(N,p)$ for each parameter set following the procedure described in Subsection~\ref{subsec:der-network-construction}; any non-strongly-connected realization is discarded.
For each valid realization of a directed Erd\H{o}s--R\'{e}nyi network, we construct the transition probability matrices of CRW and RSRW according to Eqs.~\eqref{CRW-p} and \eqref{RSRW-p}, respectively. Subsequently, we compute the MFPT $T_{ji}$ from $j$ to $i$ by applying the method detailed in Subsection~\ref{subsec:method_MFPT_MTT}. Finally, for each target node $i$, the mean trapping time $\mathrm{MTT}_i$ is calculated according to Eq.~(\ref{mtt}). This process is repeated until $\mathrm{MTT}_i$ has been computed for every target node $i$ across all valid network realizations.

To mitigate random fluctuations,
we average $\mathrm{MTT}_i$ over nodes sharing the same $d_{\mathrm{out}}(i)$
(or $d_{\mathrm{in}}(i)$) across all realizations,
and plot the averaged values against $d_{\mathrm{out}}(i)$
(or $d_{\mathrm{in}}(i)$)  to reveal the underlying relationship.
}
%To mitigate random fluctuations arising from individual network instances,  we then average $\mathrm{MTT}_i$ over nodes sharing the same $d_{\mathrm{out}}(i)$ (or $d_{\mathrm{in}}(i)$) across all valid network realizations, and plot the averaged $\mathrm{MTT}_i$ against $d_{\mathrm{out}}(i)$ (or $d_{\mathrm{in}}(i)$) to expose the dependence.

%For each valid realization of a directed Erd\H{o}s--R\'{e}nyi network, we evaluate $\mathrm{MTT}_i$ by applying the method detailed in Subsection~\ref{subsec:method_MFPT_MTT}.  This process is repeated until $\mathrm{MTT}_i$ has been computed for every target node $i$ across all valid network realizations.
%To investigate the relationship between $\mathrm{MTT}_i$ and the node degree $d_{\mathrm{out}}(i)$ (or $d_{\mathrm{in}}(i)$), we compute the average of $\mathrm{MTT}_i$ over all nodes sharing the same $d_{\mathrm{out}}(i)$ (or $d_{\mathrm{in}}(i)$) across all valid network realizations, and then plot $\mathrm{MTT}_i$ against $d_{\mathrm{out}}(i)$ (or $d_{\mathrm{in}}(i)$) to visualize their relationship.

\begin{figure}[tb]
    \centering
    \includegraphics[width=\linewidth]{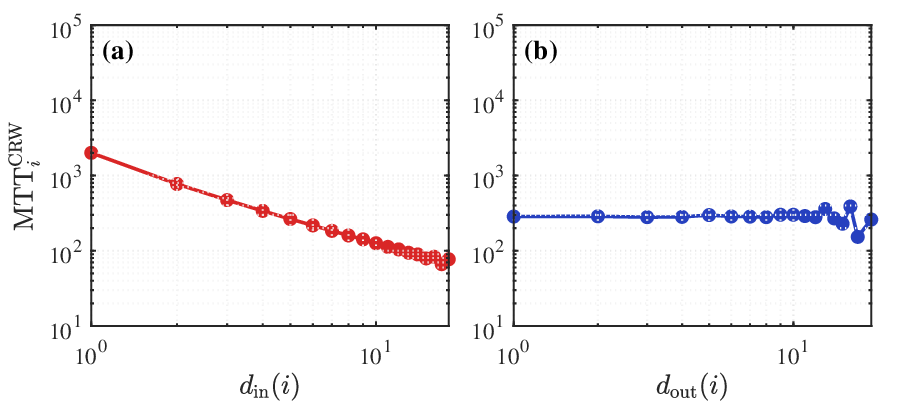}
    \caption{The averaged $\mathrm{MTT}_i^{\mathrm{CRW}}$ is shown as a function of target-node degree for CRW performed on  directed Erd\H{o}s--R\'{e}nyi networks with $N=200$ and $p=0.03$. Panels (a) and (b) correspond to $d_{\mathrm{in}}(i)$ and $d_{\mathrm{out}}(i)$, respectively. }%The red and blue curves show the degree-wise average $\mathrm{MTT}_i^{\mathrm{CRW}}$ as functions of $d_{\mathrm{in}}(i)$ and $d_{\mathrm{out}}(i)$, respectively, across all valid network realizations.
    \label{fig:DER_MTT_CRW_degree}
\end{figure}

\begin{figure}[tb]
    \centering
    \includegraphics[width=\linewidth]{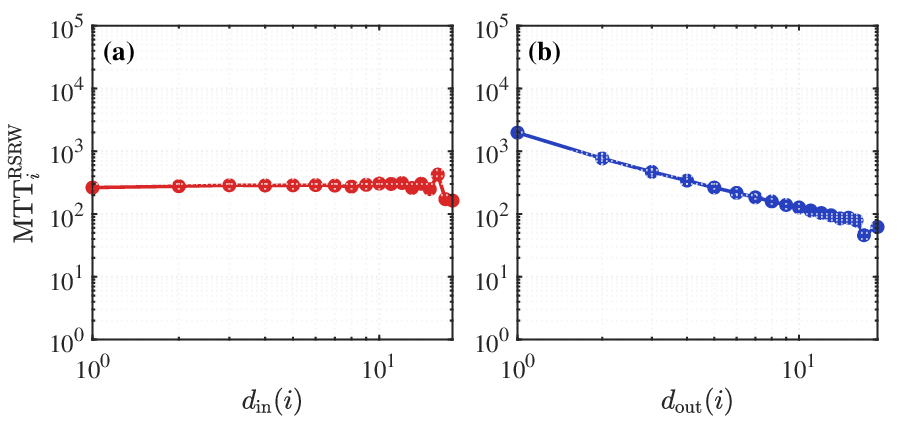}
    \caption{The averaged  $\mathrm{MTT}_i^{\mathrm{RSRW}}$ is shown as a function of target-node degree for RSRW  performed on  directed Erd\H{o}s--R\'{e}nyi networks with $N=200$ and $p=0.03$. Panels (a) and (b) correspond to $d_{\mathrm{in}}(i)$ and $d_{\mathrm{out}}(i)$, respectively.  }%The red (blue)  dot-dashed lines represent the average $\mathrm{MTT}_i^{\mathrm{RSRW}}$ over all nodes sharing the same $d_{\mathrm{out}}(i)$ (or $d_{\mathrm{in}}(i)$) across all valid network realizations.
    \label{fig:DER_MTT_RSRW_degree}
\end{figure}

Fig.~\ref{fig:DER_MTT_CRW_degree} shows the dependence of $\mathrm{MTT}_i^{\mathrm{CRW}}$ on the out-degree and in-degree of the target node under CRW.
As shown, $\mathrm{MTT}_i^{\mathrm{CRW}}$ decreases markedly with increasing in-degree $d_{\mathrm{in}}(i)$, whereas it exhibits only a weak dependence on the out-degree $d_{\mathrm{out}}(i)$.
These results indicate that, for CRW, the mean trapping time of a node is governed primarily by its in-degree, while the influence of out-degree is comparatively limited.  Fig.~\ref{fig:DER_MTT_RSRW_degree} shows the dependence of $\mathrm{MTT}_i^{\mathrm{RSRW}}$ on the out-degree and in-degree of the target node. Unlike the case of CRW, the averaged $\mathrm{MTT}_i^{\mathrm{RSRW}}$ exhibits no pronounced monotonic trend with $d_{\mathrm{in}}(i)$, but displays a clear downward trend with increasing $d_{\mathrm{out}}(i)$, indicating that nodes with larger out-degrees are trapped more efficiently under RSRW.

%This behavior can be understood from the transition rule of CRW together with the weak structural heterogeneity of directed Erd\H{o}s--R\'{e}nyi networks. Under CRW, a walker at node $u$ moves to one of its outgoing neighbors $v$ with probability $p_{uv}^{\mathrm{CRW}}=A_{uv}/d_u^{\mathrm{out}}$, as defined in Eq.~\eqref{CRW-p}. For CRW, the accessibility of a target node is mainly determined by the probability flow entering that node. For a target node $i$, this inflow comes from its predecessor nodes. In a directed Erd\H{o}s--R\'{e}nyi network, edges are generated independently and degree heterogeneity is relatively weak, so the incoming contributions from different predecessors are comparable on average. The total inflow to node $i$ is therefore governed mainly by its in-degree $d_{\mathrm{in}}(i)$. This explains why nodes with larger in-degree are reached more easily and have smaller $\mathrm{MTT}_i^{\mathrm{CRW}}$. By contrast, the out-degree $d_{\mathrm{out}}(i)$ mainly affects the motion after arrival, and thus has a comparatively weak effect on $\mathrm{MTT}_i^{\mathrm{CRW}}$.

%In other words, while CRW is governed primarily by in-degree, RSRW is shaped predominantly by out-degree.
Therefore, on directed Erd\H{o}s--R\'{e}nyi networks,
CRW yields smaller MTTs at nodes with high in-degree,
whereas RSRW achieves smaller MTTs at nodes with high out-degree.
Consequently, CRW tends to more readily visit nodes with large in-degree,
while RSRW is more effective at locating nodes with large out-degree.

\section{Random walks on directed scale-free networks}
\label{sec:DSF}

In this section, we first introduce a novel directed network model—the directed scale-free network—and use it as a case study.
We begin by numerically validating that the stationary distribution, mean trapping time (MTT), and Kemeny's constant for RSRW on the original directed graph~$G$ coincide with those for CRW on the reverse graph~$G^{r}$ across various network configurations.
Subsequently, we examine how the MTT depends on the in-degree and out-degree of nodes for both CRW and RSRW on~$G$.
%In this section, we first introduce a novel directed network model: the directed scale-free network. We then analyze the dynamics of both the CRW and the RSRW performed on this network. Our analysis focuses on comparing the stationary distributions and the mean trapping time (MTT) of the two random walk mechanisms, with particular emphasis on the impact of in- and out-degrees on the MTT.

\subsection{Network construction}
\label{subsec:DSF-network}
The directed scale-free networks, considered here,  start from a simple strongly connected  directed graph (e.g., a directed ring with $3$ nodes), and grow iteratively via preferential attachment. Let $G(t)=(V(t),E(t))$ denote the network at time $t$, and let $d_i^{\mathrm{in}}\left(t\right)$ and $d_i^{\mathrm{out}}\left(t\right)$ be the in- and out-degrees of node $i$ at time $t$, respectively.  The probabilities of attaching to an existing node as the target of an incoming edge or as the source of an outgoing edge  are given by
\begin{align}
\Pi_u^{\mathrm{in}}(t)% ($\Pi_u^{\mathrm{in}}$)
&=
\frac{d_u^{\mathrm{in}}(t)+\delta_{\mathrm{in}}}
{\sum_{v\in V(t)}\left[d_v^{\mathrm{in}}(t)+\delta_{\mathrm{in}}\right]},
\label{eq:preferential-attachment-in}\\
\Pi_u^{\mathrm{out}}(t) %($\Pi_u^{\mathrm{out}}$)
&=
\frac{d_u^{\mathrm{out}}(t)+\delta_{\mathrm{out}}}
{\sum_{v\in V(t)}\left[d_v^{\mathrm{out}}(t)+\delta_{\mathrm{out}}\right]},
\label{eq:preferential-attachment-out}
\end{align}
respectively, where $\delta_{\mathrm{in}}\geq 0$ and $\delta_{\mathrm{out}}\geq 0$ are parameters governing the strength of preferential attachment.

%At each discrete time step, the network evolves as follows. With probability $\mu\in(0,1)$, a new node $v$ is introduced. An existing node $u$ is selected with probability $\Pi_u^{\mathrm{out}}\left(t\right)$, and form a directed edge from  $u$ to $v$; another existing node $w$ is selected with probability $\Pi_w^{\mathrm{in}}\left(t\right)$, and form a directed edge from $v$ to $w$. Thus each new node enters with one incoming and one outgoing edge. With probability $1-\mu$, an edge is added between existing nodes: the initial node $u$ is sampled according to $\Pi_i^{\mathrm{out}}\left(t\right)$, the terminal node $w$ according to $\Pi_i^{\mathrm{in}}\left(t\right)$, and the edge $u\rightarrow w$ is added.

%At each step $t$, the network evolves as follows: with probability $\mu \in (0,1)$, a new node $v$ is introduced. An existing node $u$ is selected with probability $\Pi_u^{\mathrm{out}}(t)$ and \textbf{forms} a directed edge from $u$ to $v$; another existing node $w$ is selected with probability $\Pi_w^{\mathrm{in}}(t)$ and \textbf{forms} a directed edge from $v$ to $w$. Thus, each new node enters with one incoming and one outgoing edge. With probability $1-\mu$, an edge is added between existing nodes: the initial node $u$ is selected according to $\Pi_i^{\mathrm{out}}(t)$, the terminal node $w$ according to $\Pi_i^{\mathrm{in}}(t)$, and a directed edge from $u$ to $w$ is added.

At each step $t$, the network evolves as follows. \\
{(i) Node addition ($\mu$):} With probability $\mu \in (0,1)$, a new node $v$ is introduced. We select an existing node $u$ with probability $\Pi_u^{\mathrm{out}}(t)$ and create a directed edge $u \to v$. Simultaneously, we select another existing node $w$ with probability $\Pi_w^{\mathrm{in}}(t)$ and create a directed edge $v \to w$. Hence, every new node enters the network with exactly one incoming and one outgoing edge. \\
{(ii) Edge addition ($1-\mu$):} With probability $1-\mu$, a directed edge is added between existing nodes. The source node $u$ is chosen with probability $\Pi_u^{\mathrm{out}}(t)$, and the target node $w$ is chosen with probability $\Pi_w^{\mathrm{in}}(t)$, forming the directed edge $u \to w$.
%Therefore, these networks are parameterized by three parameters—$\mu \in (0,1)$, $\delta_{\mathrm{in}}\geq 0$, and $\delta_{\mathrm{out}}\geq 0$, where $\mu$ control the probability a new node is introduced at each step, and  $\delta_{\mathrm{in}}$ and $\delta_{\mathrm{out}}$ control the strength of preferential attachment of node degree.
Therefore, the networks are characterized by three parameters: $\mu \in (0,1)$, $\delta_{\mathrm{in}}\geq 0$, and $\delta_{\mathrm{out}}\geq 0$. Here, $\mu$ controls the probability of adding a new node at each time step, while $\delta_{\mathrm{in}}$ and $\delta_{\mathrm{out}}$ determine the strength of preferential attachment with respect to in-degree and out-degree, respectively.

Unlike the directed scale-free growth model of Bollobás et al.~\cite{Bollobas-2003-DSFG}, the present construction maintains strong connectivity throughout the growth process and permits independent tuning of the in- and out-degree distributions via the preference parameters $\delta_{\mathrm{in}}$ and $\delta_{\mathrm{out}}$.
Critically, the original model does not generally guarantee strong connectivity.
To address this, we replace the original two-step node-addition process with a unified event:
with probability $\mu$, a new vertex $v$ is introduced and simultaneously attaches to the existing network via two directed edges, $v \to u$ and $w \to v$.
Here, the preexisting vertices $u$ and $w$ are sampled according to the in-degree and out-degree preferential probabilities, $\Pi_u^{\mathrm{in}}$ and $\Pi_w^{\mathrm{out}}$, respectively.
Consequently, provided the initial seed graph is strongly connected, the evolving network remains strongly connected by construction.

%However,  the original model presented in Ref.~\cite{Bollobas-2003-DSFG} do not, in general, guarantee that the generated network is strongly connected. Here, the two alternative node-addition events in the original model are replaced here by a single node-addition event. In the original model, a new vertex is added either as the source of an edge pointing to an existing vertex, with probability $\alpha$, or as the target of an edge from an existing vertex, with probability $\gamma$. In the present model, with probability $\mu$, a new vertex $v$ is introduced, and two directed edges, $v\to u$ and $w\to v$, are added simultaneously. Here the existing vertex $u$ is selected according to the in-degree preferential probability $\Pi_u^{\mathrm{in}}$, while the existing vertex $w$ is selected according to the out-degree preferential probability $\Pi_w^{\mathrm{out}}$. As a result, provided that the initial seed graph is strongly connected, the generated network remains strongly connected throughout the growth process.

\subsection{Degree distribution}
%偏好参数控制中等规模网络出入度的分布
As shown in Appendix~\ref{appendix:Asymptotic-Degree-Distribution}, for networks generated by the present model in the asymptotic large-network limit, the out-degree and in-degree distributions admit the asymptotic forms
\begin{equation}
P_{\mathrm{out}}(k)
\sim
k^{-\left[2+\mu(1+\delta_{\mathrm{out}})\right]} ,
\label{eq:out-degree-exponent}
\end{equation}
and
\begin{equation}
P_{\mathrm{in}}(k)
\sim
k^{-\left[2+\mu(1+\delta_{\mathrm{in}})\right]} ,
\label{eq:in-degree-exponent}
\end{equation}
respectively. Thus, both distributions exhibit power-law scaling in their tails in this limit.

However, for moderate-size networks, the parameters $\delta_{\mathrm{in}}$ and $\delta_{\mathrm{out}}$ allow us to tune the in-degree and out-degree distributions between power-law tails and exponential decay.

For instance, when $\delta_{\mathrm{in}} \gg d_{\mathrm{in}}(i)$, Eq.~\eqref{eq:preferential-attachment-in} reduces to an approximately uniform attachment probability, as $d_{\mathrm{in}}(i)+\delta_{\mathrm{in}} \approx \delta_{\mathrm{in}}$. Consequently,
\begin{equation}
\Pi_i^{\mathrm{in}}(t) \approx \frac{1}{N(t)} ,
\label{eq:preferential-attachment-in-uniform}
\end{equation}
rendering the growth process nearly independent of the current in-degree. The resulting in-degree distribution consequently becomes significantly narrower, typically exhibiting exponential-like decay. Conversely, when $d_{\mathrm{in}}(i) \gg \delta_{\mathrm{in}}$, one has $d_{\mathrm{in}}(i)+\delta_{\mathrm{in}} \approx d_{\mathrm{in}}(i)$, and the attachment probability becomes dominated by the in-degree. This recovers the standard preferential-attachment regime, yielding a power-law distribution. An analogous argument applies to the out-degree distribution, which is governed by $\delta_{\mathrm{out}}$.

%文中分4类说明中等规模网络中，度分布出现4中不同清晰。列出4类。重点用图解释一类，其他几类简单说明一下类同。若图表达简单的话可以也显示出来。

%We use the complementary cumulative distribution function (CCDF) to examine the decay forms of the out-degree and in-degree distributions. If the degree distribution follows a power-law decay, the CCDF appears approximately as a straight line in double-logarithmic coordinates, i.e., $\log(\mathrm{degree})$ versus $\log P(D \geq \mathrm{degree})$. By contrast, if the degree distribution is close to exponential decay, the CCDF appears approximately linear in semi-logarithmic coordinates, i.e., $\mathrm{degree}$ versus $\log P(D \geq \mathrm{degree})$. For different node-addition probabilities $\mu$ and different parameter sets $(\delta_{\mathrm{in}}, \delta_{\mathrm{out}})$, we vary only the network size $N$ and examine how the degree distributions behave through the relation between degree and $\log P(D \geq \mathrm{degree})$.

%delta_in=10，delta_out=1
\begin{figure}[tb]
    \centering
    \includegraphics[width=\columnwidth]{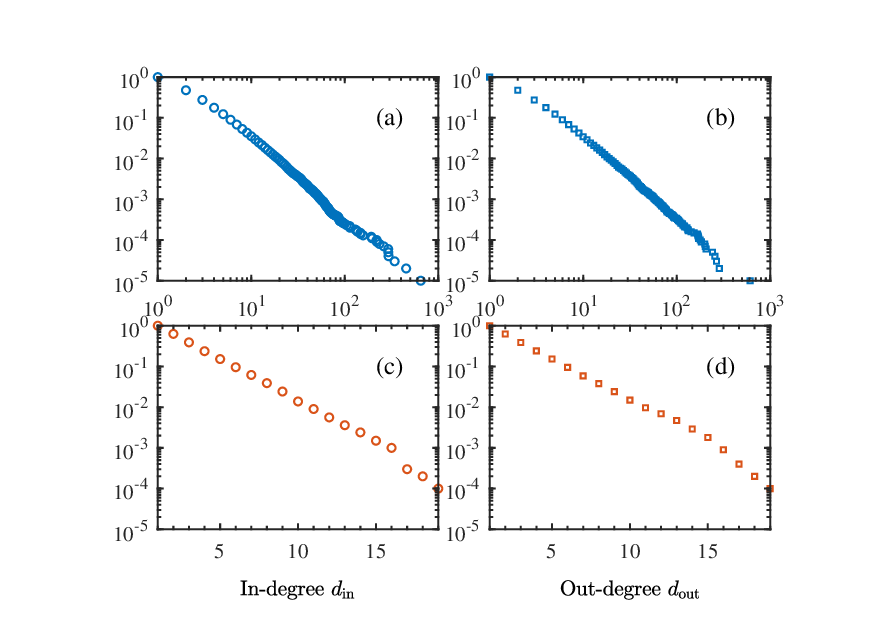}
    \caption{Cumulative degree distributions of directed scale-free networks of moderate size and large size with $\mu=0.6$.
    Panels (a,b) show the in-degree and out-degree distributions, respectively, for the large network with
    $N=1\times10^{5}$ and $\delta_{\mathrm{in}}=\delta_{\mathrm{out}}=1$ in double-logarithmic coordinates.
    Panels (c,d) show the corresponding distributions for the moderate-size network with
    $N=10^{4}$ and $\delta_{\mathrm{in}}=\delta_{\mathrm{out}}=100$ in semi-logarithmic coordinates.
}
\label{fig:DSF-CCDF-delta}
\end{figure}
%FIG.~\ref{fig:DSF-CCDF-delta} illustrates how the parameters $\delta_{\mathrm{in}}$ and $\delta_{\mathrm{out}}$ affect the cumulative degree distributions (CDFs) of the modified directed scale-free networks. For $\delta_{\mathrm{in}}=\delta_{\mathrm{out}}=1$ and a large network size $N=10^5$, both the in-degree and out-degree CDFs are approximately linear in double-logarithmic coordinates [FIGs.~\ref{fig:DSF-CCDF-delta}(a) and \ref{fig:DSF-CCDF-delta}(b)], indicating power-law decay. In contrast, for $\delta_{\mathrm{in}}=\delta_{\mathrm{out}}=100$ and a moderate network size $N=10^4$, the corresponding CDFs are nearly linear in semi-logarithmic coordinates [FIGs.~\ref{fig:DSF-CCDF-delta}(c) and \ref{fig:DSF-CCDF-delta}(d)], showing exponential-like decay over the observed range.

%This finite-size behavior reflects the role of $\delta_{\mathrm{in}}$ and $\delta_{\mathrm{out}}$ in the preferential-attachment kernels: small values of the parameters make attachment primarily degree-driven and enhance the rich-get-richer effect, whereas large values weaken the degree dependence over the accessible degree range and make the attachment process closer to uniform. Nevertheless, this exponential-like behavior is a finite-size effect; in the asymptotic limit $N\to\infty$, the degree distributions generated by the model still have power-law tails.

{
Fig.~\ref{fig:DSF-CCDF-delta} shows the effect of $\delta_{\mathrm{in/out}}$ on cumulative degree distributions (CDFs) of the directed scale-free networks: $\delta_{\mathrm{in}}=\delta_{\mathrm{out}}=1$, $N=10^5$: double-log-linear CDFs [Figs.~\ref{fig:DSF-CCDF-delta}(a,b)] confirm power-law decay; $\delta_{\mathrm{in}}=\delta_{\mathrm{out}}=100$, $N=10^4$: semi-log-linear CDFs [Figs.~\ref{fig:DSF-CCDF-delta}(c,d)] indicate exponential decay.

This finite-size effect arises because small $\delta_{\mathrm{in/out}}$ strengthen degree-driven preferential attachment (rich-get-richer), while large $\delta_{\mathrm{in/out}}$ suppress degree dependence, yielding near-uniform attachment. The exponential-like decay is a finite-size artifact; power-law tails persist asymptotically ($N\to\infty$).

We adopt a generate-first, clean-later workflow: candidate edges are added per growth rules without self-loop/duplicate checks, then purged post-growth to yield the simple directed graph $\hat{G}$. This causes minor, transient degree perturbations and slight edge-count reduction; in the large-$N$ regime considered, these biases are negligible. All subsequent MFPT/MTT analyses use postprocessed $\hat{G}$, with only strongly connected simple graphs retained for verification of the stationary distribution, MTT, and Kemeny’s constant.
}

%The final network used in the random-walk analysis is a simple directed graph. For computational efficiency, however, we adopt a ``generate first, clean later'' implementation: during growth, candidate edges are recorded directly according to the above rules, without checking for self-loops or repeated edges. After the prescribed size is reached, all self-loops and duplicate edges are removed in one postprocessing step, producing the simple directed graph $\hat{G}$.

%This procedure allows edges that are later removed to contribute temporarily to degree accumulation, slightly perturbing subsequent attachment probabilities; it also makes the final number of edges marginally smaller than the number of generated candidates. In the large-network regime considered here, self-loops and multiple edges occur with low frequency, so the induced bias is treated as a local correction that does not affect the feasibility of the numerical MFPT and MTT analyses. The postprocessed graph $\hat{G}$ is therefore used as the network sample in what follows. {Only network samples that remain simple and strongly connected directed graphs after postprocessing are retained for the subsequent numerical verification of the stationary distribution, mean trapping time, and Kemeny's constant.}
%\textcolor{red}Only samples that remain strongly connected after postprocessing are retained. The modified directed scale-free network construction is thus fully specified, and its topological properties are examined in the next subsection.

\subsection{Equivalence of stationary and transient properties: RSRW on $G$ vs. CRW on $G^r$}
\label{subsec:DSF_verification}
{

In this subsection, following the same method as for directed Erd\H{o}s--R\'{e}nyi networks (see Sec.~\ref{sec:random-walks-directed-er}), we verify the theoretical equivalence of key properties—i.e., the stationary distribution, mean trapping time ($\mathrm{MTT}$), and Kemeny's constant—between RSRW on the original graph $G$ and CRW on the reverse graph $G^r$, as established in Sec.~\ref{subsec:Comparison_RSRW_CRW}.

Numerical results confirm that across all tested parameter regimes, the stationary distribution, $\mathrm{MTT}$, and Kemeny's constant of RSRW on $G$ differ from those of CRW on $G$, but coincide exactly with the corresponding metrics of CRW on $G^r$. Detailed empirical evidence supporting the equivalence of the stationary distributions is provided in Appendix~\ref{appendix:DSF-stationary-distribution}.

}

%\subsection{Effect of in-degree (or out degree) on the MTT for both CRW and RSRW strategies}
%\label{subsec:DSF-mtt}
%\subsection{Effect of In-Degree and Out-Degree on the MTT for CRW and RSRW}
%\label{subsec:degree_effect-SF}
%\subsection{Mean trapping time and Kemeny's const}
%\label{subsec:degree_effect-SF}
\subsection{Degree dependence of mean trapping times}
\label{subsec:DSF_degree_dependence}
In this subsection, we investigate how the MTT depends on the in-degree and out-degree of nodes for both CRW and RSRW on directed scale-free networks.

{ Following the same method as for the directed Erd\H{o}s--R\'{e}nyi networks (Sec.~\ref{subsec:ER_degree_dependence}),
we generate 100 valid scale-free realizations for each parameter set $(N, \mu, \delta_{\mathrm{in}}, \delta_{\mathrm{out}})$
via the construction in Sec.~\ref{subsec:DSF-network}.
For each realization, we compute $\mathrm{MTT}_i$ for every target node $i$
using the method in Subsec.~\ref{subsec:method_MFPT_MTT}.
To suppress statistical fluctuations inherent to finite-size networks,
we take the median of $\mathrm{MTT}_i$ over nodes sharing the same $d_{\mathrm{out}}(i)$
(resp.\ $d_{\mathrm{in}}(i)$) pooled across all realizations.
We then plot this median against the corresponding degree to characterize the underlying relationship.
Unlike the homogeneous ER case where the mean suffices, we adopt the median here because scale-free networks exhibit heavy-tailed MTT distributions; the median reduces the influence of extreme values and thus provides a more robust characterization of the typical degree dependence than the mean.}

%To characterize the degree dependence of $\mathrm{MTT}_i$, nodes with the same $d_{\mathrm{in}}(i)$ or $d_{\mathrm{out}}(i)$ are pooled over all valid realizations, and the median $\mathrm{MTT}_i$ is plotted against the corresponding degree. The median is used to reduce the influence of extreme values within the same degree class.

\begin{figure}[tb]
\centering
\includegraphics[width=\columnwidth]{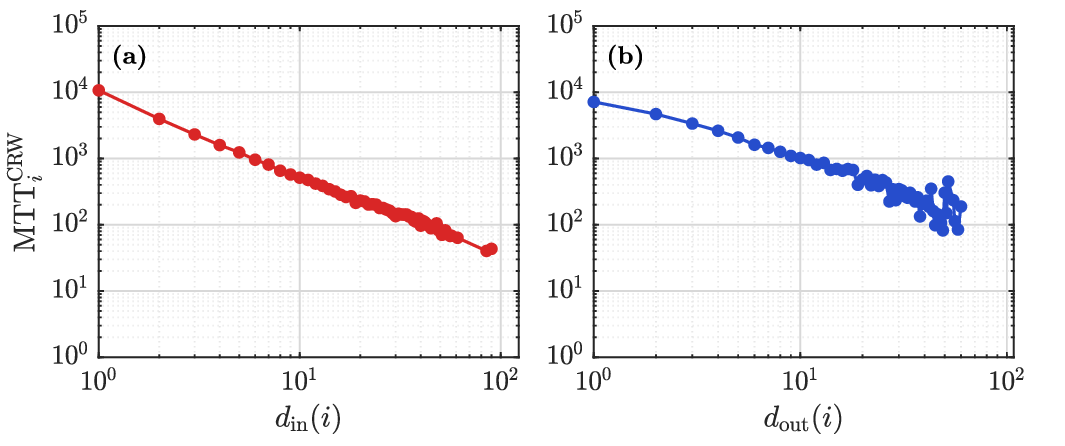}
\caption{The median $\mathrm{MTT}_i^{\mathrm{CRW}}$ is plotted as a function of target-node degree for CRW on directed scale-free networks with $N=2000$, $\mu=0.6$, and $\delta_{\mathrm{in}}=\delta_{\mathrm{out}}=1$.
Panels (a) and (b) display the results for $d_{\mathrm{in}}(i)$ and $d_{\mathrm{out}}(i)$, respectively.
The red (blue) solid curves represent the median $\mathrm{MTT}_i^{\mathrm{CRW}}$ computed over nodes sharing the same $d_{\mathrm{in}}(i)$ (resp.\ $d_{\mathrm{out}}(i)$) across all valid network realizations.
Both axes are scaled logarithmically.}
%\caption{Dependence of the $\mathrm{MTT}_i^{\mathrm{CRW}}$ on local degree in directed scale-free networks. Panels (a) and (b) show $\mathrm{MTT}_i^{\mathrm{CRW}}$ as functions of the in-degree $d_{\mathrm{in}}\left(i\right)$ and out-degree $d_{\mathrm{out}}\left(i\right)$, respectively. Gray markers denote pooled node-level data, and solid curves indicate median trends. Both axes are shown on logarithmic scales.}
\label{fig:DSF-classical-MTT-degree}
\end{figure}

Fig.~\ref{fig:DSF-classical-MTT-degree} shows the dependence of $\mathrm{MTT}_i$ on the out-degree and in-degree of the target node under CRW. In both panels, the  plots show a systematic decrease of $\mathrm{MTT}_i$ as either $d_{\mathrm{out}}(i)$ or $d_{\mathrm{in}}(i)$ increases.
Therefore, in directed scale-free networks, the dependence of $\mathrm{MTT}_i$ on out-degree and in-degree appears similar, so the dominant degree factor affecting CRW cannot be clearly identified from this comparison alone.

\begin{figure}[tb]
\centering
\includegraphics[width=\columnwidth]{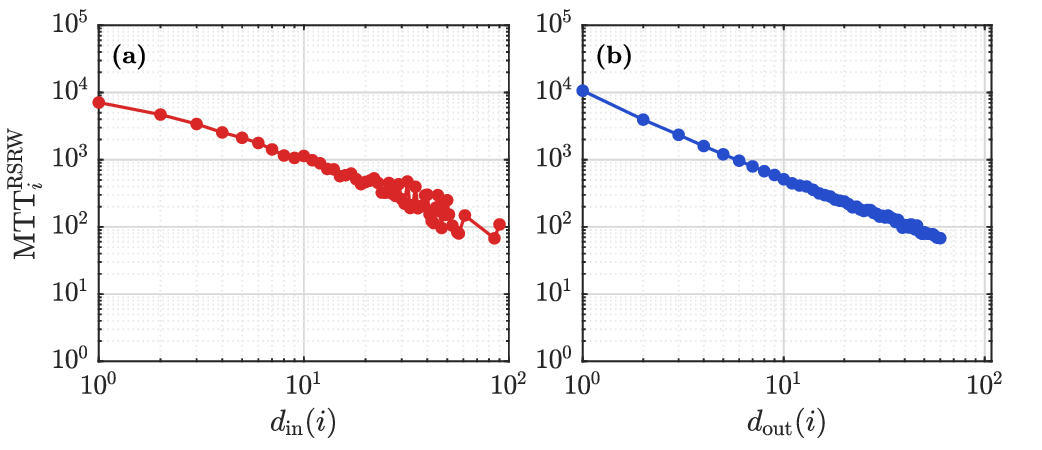}
\caption{The median $\mathrm{MTT}_i^{\mathrm{RSRW}}$ is plotted as a function of target-node degree for RSRW on directed scale-free networks with $N=2000$, $\mu=0.6$, and $\delta_{\mathrm{in}}=\delta_{\mathrm{out}}=1$.
Panels (a) and (b) display the results for $d_{\mathrm{in}}(i)$ and $d_{\mathrm{out}}(i)$, respectively.
The red (blue) solid curves represent the median $\mathrm{MTT}_i^{\mathrm{RSRW}}$ computed over nodes sharing the same $d_{\mathrm{in}}(i)$ (resp.\ $d_{\mathrm{out}}(i)$) across all valid network realizations.
Both axes are scaled logarithmically.}
%\caption{Dependence of $\mathrm{MTT}_i^{\mathrm{RSRW}}$ under the RSRW mechanism on node degree in directed scale-free networks. Panels (a) and (b) show the dependence on $d_{\mathrm{in}}\left(i\right)$ and $d_{\mathrm{out}}\left(i\right)$, respectively. Gray markers denote pooled node-level data, and solid curves show median trends.}
\label{fig:DSF-new-MTT-degree}
\end{figure}
Fig.~\ref{fig:DSF-new-MTT-degree} presents the corresponding results for the RSRW mechanism. Similar to the CRW case, $\mathrm{MTT}_i$ decreases systematically with increasing $d_{\mathrm{out}}(i)$ and also with increasing $d_{\mathrm{in}}(i)$. These overall degree-based plots do not allow us to directly identify which degree variable plays the dominant role in determining $\mathrm{MTT}_i$ under RSRW. %Therefore, a further analysis is needed to distinguish the respective contributions of the in-degree and the out-degree.

\begin{figure}[tb]
\centering
\includegraphics[width=\columnwidth]{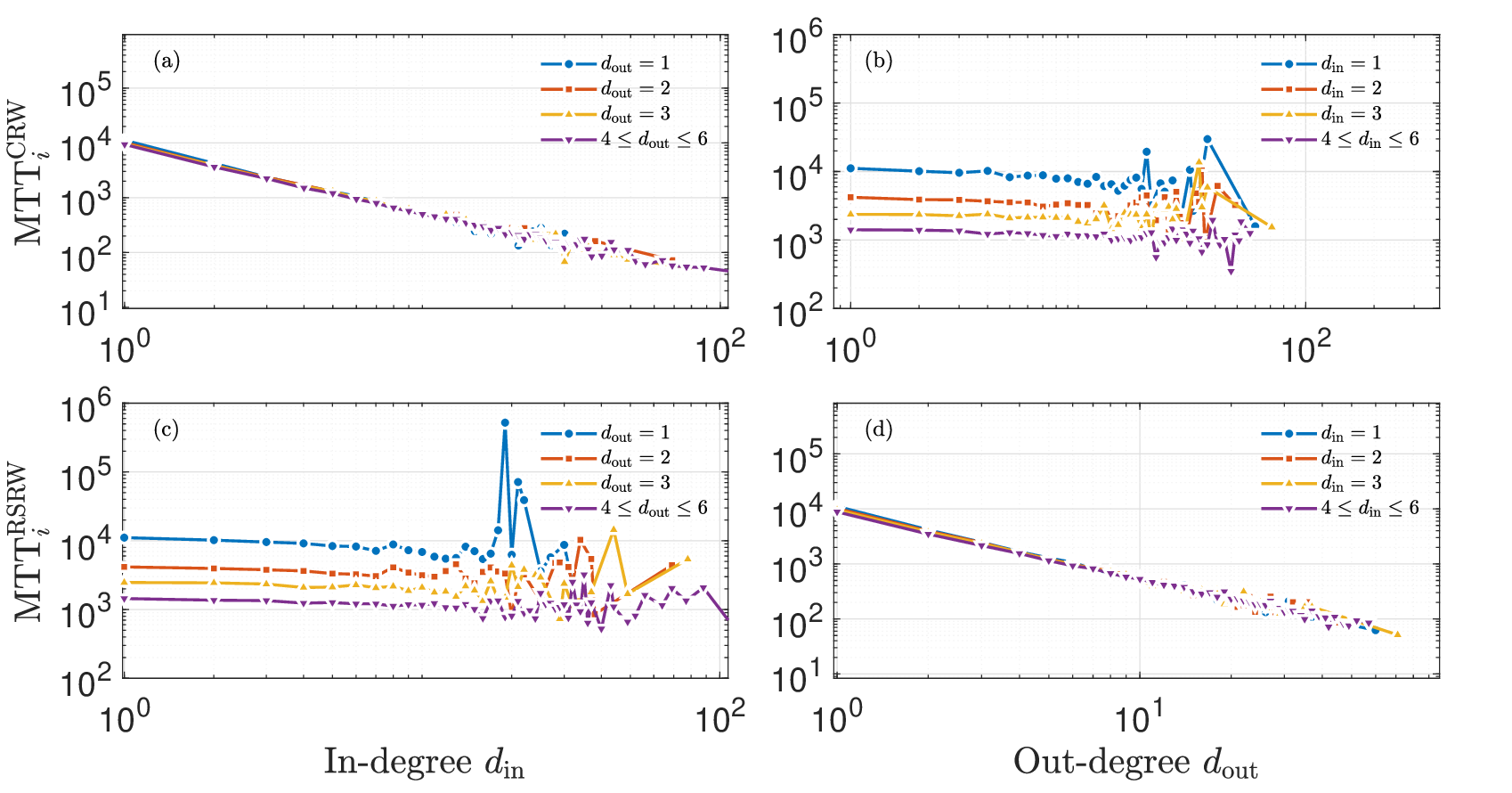}
\caption{Relation between the median $\mathrm{MTT}_i$ and in-degree (or out-degree) for nodes sharing similar or identical out-degrees (or in-degrees). Panels (a) and (b) show the median $\mathrm{MTT}_i^{\mathrm{CRW}}$ as a function of in-degree and out-degree, respectively, while panels (c) and (d) show the corresponding results for $\mathrm{MTT}_i^{\mathrm{RSRW}}$. The network parameters are set to $N=2000$, $\mu=0.6$, and $\delta_{\mathrm{in}}=\delta_{\mathrm{out}}=1$.
Statistics are collected over $100$ independent realizations, where $\mathrm{MTT}_i$ represents the median value across all nodes sharing the same integer degree. The pronounced tail fluctuations arise from insufficient sampling of high-degree nodes, leading to statistical uncertainty.}
\label{fig:MTT_degree_layers_2x2}
\end{figure}

{ This phenomenon stems from the strong positive correlation between in- and out-degrees in directed scale-free networks. In networks grown via preferential attachment, early nodes accumulate both high in- and out-degrees, while newer nodes exhibit lower values for both. This coupling entangles the effects of $d_{\mathrm{in}}(i)$ and $d_{\mathrm{out}}(i)$, obscuring the identification of the dominant degree factor—--unlike the decoupled regime observed in directed Erd\H{o}s--R\'{e}nyi networks.

To further elucidate the degree dependence of $\mathrm{MTT}_i$, we analyze its response to one degree measure while holding the other fixed. Specifically, to isolate the effect of out-degree, we pool data across all 100 valid network realizations, and categorize nodes into groups based on their in-degree, such that nodes within the same group share similar or identical in-degrees. %Nodes with identical or similar $d_{\mathrm{in}}(i)$ are binned into groups.
Within each group, we compute the median $\mathrm{MTT}_i$ across nodes sharing the same $d_{\mathrm{out}}(i)$, and subsequently characterize the dependence of this median on $d_{\mathrm{out}}(i)$.
Due to the scarcity of samples for high in-degree nodes, which hinders the isolation of out-degree effects, we restrict our analysis to nodes with lower in-degrees. For instance, in networks of size $N=2000$, nodes are partitioned into four in-degree groups: $d_{\mathrm{in}}=1$, $d_{\mathrm{in}}=2$, $d_{\mathrm{in}}=3$, and $4 \le d_{\mathrm{in}} \le 6$.
Nodes with other in-degree values are excluded from this layer-specific analysis due to insufficient sample sizes.
Fig.~\ref{fig:MTT_degree_layers_2x2}(a) and (c) illustrate the relationship between the median $\mathrm{MTT}_i$ and out-degree for CRW and RSRW, respectively, for nodes sharing  similar or identical in-degrees on networks with parameters $N=2000$, $\mu=0.6$, and $\delta_{\mathrm{in}}=\delta_{\mathrm{out}}=1$.
Similarly, we also examine the influence of in-degree on the $\mathrm{MTT}_i$ for nodes sharing similar or identical out-degrees. Fig.~\ref{fig:MTT_degree_layers_2x2}(b) and (d) illustrate the relationship between $\mathrm{MTT}_i$ and in-degree for CRW and RSRW, respectively, under the constraint of fixed or similar out-degree.
}

%Fig.~\ref{fig:MTT_degree_layers_2x2} further resolves the dependence of $\mathrm{MTT}_i$ on in-degree and out-degree under fixed-degree conditioning.
One can find, for the CRW dynamics [Fig.~\ref{fig:MTT_degree_layers_2x2}(a) and (b)], $\mathrm{MTT}_i^{\mathrm{CRW}}$ decreases with increasing $d_{\mathrm{in}}$ when nodes are grouped by fixed $d_{\mathrm{out}}$, with no clear separation among different out-degree groups.
In contrast, when nodes are grouped by fixed $d_{\mathrm{in}}$, $\mathrm{MTT}_i^{\mathrm{CRW}}$ remains invariant with respect to the out-degree, and the curves remain well separated by the in-degree layers, and larger $d_{\mathrm{in}}$ classes yield smaller $\mathrm{MTT}_i^{\mathrm{CRW}}$.
These findings indicate that the out-degree has virtually no effect on $\mathrm{MTT}_i^{\mathrm{CRW}}$, whereas the in-degree exerts a substantial influence; specifically, nodes with larger in-degrees exhibit significantly smaller $\mathrm{MTT}_i^{\mathrm{CRW}}$. For the RSRW dynamics [Fig.~\ref{fig:MTT_degree_layers_2x2}(c) and (d)], the opposite trend is observed and indicates that the in-degree has virtually no effect on $\mathrm{MTT}_i^{\mathrm{RSRW}}$, whereas the out-degree exerts a substantial influence; specifically, nodes with larger out-degrees exhibit significantly smaller $\mathrm{MTT}_i^{\mathrm{RSRW}}$.

{
\section{Random Walks on an ArXiv citation network} \label{sec:cite_network}
%In this section, we use the real-world Cit-HepTh citation network as a case study. We first validate in the Cit-HepTh network the degree-dependent MTT behavior previously observed in directed ER and SF networks, and then examine how the relative trapping efficiency of the two walks varies with the out-degree-to-in-degree ratio of the target nodes. Subsequently, we compare the transient degree profiles of the nodes visited by the two walks, quantified by $\mathbb{E}[d_{\mathrm{in}}(X_t)]$ and $\mathbb{E}[d_{\mathrm{out}}(X_t)]$, to characterize their distinct exploration preferences.

In this section, we use the arXiv High Energy Physics citation network (i.e., the Cit-HepTh network)~\cite{snap-cit-hepth} as a case study.
We first validate the degree-dependent mean trapping time (MTT) behavior observed previously in directed Erd\H{o}s--R\'{e}nyi and scale-free networks within the Cit-HepTh network. Subsequently, by comparing the in- and out-degrees of nodes visited at each step by CRW and RSRW, we demonstrate distinct visit preferences with respect to nodal in-degree and out-degree.

\subsection{Dataset and network construction}

The Cit-HepTh dataset, obtained from the Stanford Large Network Dataset Collection (SNAP)~\cite{snap-cit-hepth}, is an arXiv citation network in the field of high-energy physics theory that has been analyzed in previous studies of evolving network structure~\cite{leskovec-2005-acm}. Each node represents a paper in the high-energy physics theory category, and a directed edge $i\rightarrow j$ indicates that paper $i$ cites paper $j$. Under this convention, the in-degree of a node measures the number of papers in the network that cite the corresponding paper, whereas its out-degree measures the number of references pointing to other papers retained in the network.

The Cit-HepTh network contains 27,770 nodes and 352,807 directed edges. The Spearman rank correlation between the in-degree and out-degree of the nodes is approximately 0.30. After removing self-loops and duplicated edges, we extract the largest strongly connected component (LSCC), which contains 7,464 nodes and 116,252 directed edges. In the resulting component, the Spearman correlation between the in-degree and out-degree decreases to approximately 0.18. All subsequent analyses are performed on this LSCC.

Restricting the analysis to the LSCC ensures that every retained node is reachable from every other retained node through directed paths. Therefore, CRW and RSRW are defined on the same state space, allowing their stationary distributions, mean first-passage times, and mean trapping times to be consistently calculated and compared. Notably, the weaker in--out degree correlation in the LSCC indicates that the two degree measures characterize more distinguishable structural roles. This property facilitates the examination of whether the trapping efficiencies of CRW and RSRW are associated with different degree characteristics of the target nodes. We next examine whether these distinct in- and out-degree roles are reflected in the trapping efficiencies of CRW and RSRW.

\subsection{Degree dependence of mean trapping times}

In this subsection, we use the LSCC of the Cit-HepTh network as a case study to examine whether the degree-dependent MTT behavior observed in directed ER and SF networks persists in a real-world network and how the relative trapping efficiency of RSRW and CRW on the LSCC varies with the out-degree-to-in-degree ratio of the target nodes.
%In this subsection, we use the LSCC of the Cit-HepTh network  (referred to as G for simplicity) as a case study to examine whether the degree-dependent MTT behavior observed in directed ER and SF networks persists in a real-world network, and how the relative trapping efficiency of RSRW and CRW on G varies with the out-degree-to-in-degree ratio of the target nodes.

First, we investigate how the MTT depends on the in-degree and out-degree of nodes for both CRW and RSRW on the LSCC. We calculate $\mathrm{MTT}_i$  for each target node $i$ using the method detailed in subsection \ref{subsec:method_MFPT_MTT}.

%For the LSCC of the Cit-HepTh network, we construct the transition probability matrices of CRW and RSRW according to Eqs.~\eqref{CRW-p} and \eqref{RSRW-p}, respectively. We then compute the MFPT $T_{ij}$ between every pair of nodes $i$ and $j$ using the method described in Subsection~\ref{subsec:method_MFPT_MTT}. Based on the resulting MFPT matrix, the mean trapping time $\mathrm{MTT}_j$ is evaluated for each target node $j$ according to Eq.~\eqref{mtt}.

To expose the dependence of $\mathrm{MTT}_i$ on $d_{\mathrm{in}}$ (resp.\ $d_{\mathrm{out}}$),
nodes are binned by $d_{\mathrm{in}}$ (resp.\ $d_{\mathrm{out}}$), and the median $\mathrm{MTT}_i$ is computed within each bin. The resulting median is then plotted against $d_{\mathrm{in}}$ (resp.\ $d_{\mathrm{out}}$) to characterize the degree-dependent trend.

%To examine the dependence of $\mathrm{MTT}_i$ on the in-degree (resp. out-degree) of the target node, we group the nodes according to their in-degree (resp. out-degree) such that all nodes in a group sharing the same  in-degree (resp. out-degree). In each group, we calculate the median MTT over all nodes sharing the same degree. The resulting degree-based median MTT is then plotted against $d_{\mathrm{in}}$ (or $d_{\mathrm{out}}$), respectively, to reveal the corresponding degree-dependent trends. For each distinct value of $d_{\mathrm{in}}$ (or $d_{\mathrm{out}}$), we calculate the median MTT over all nodes sharing the same degree. The resulting degree-based median MTT is then plotted against $d_{\mathrm{in}}$ (or $d_{\mathrm{out}}$), respectively, to reveal the corresponding degree-dependent trends.

\begin{figure}[tb]
    \centering
    \includegraphics[width=\linewidth]{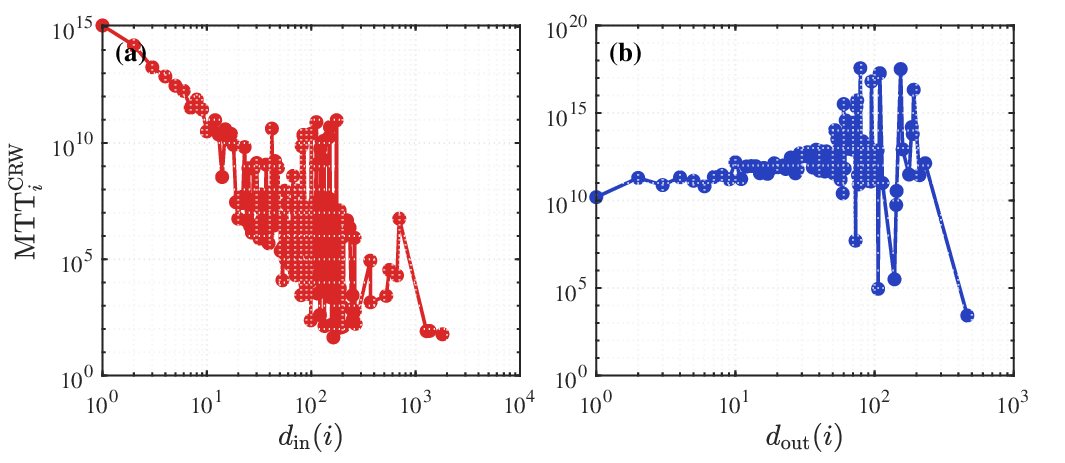}
    \caption{$\mathrm{MTT}_i^{\mathrm{CRW}}$ is shown as a function of target-node degree for CRW performed on the LSCC of the Cit-HepTh network. The red and blue curves in panels (a) and (b) show the degree-wise median $\mathrm{MTT}_i^{\mathrm{CRW}}$ as functions of $d_{\mathrm{in}}(i)$ and $d_{\mathrm{out}}(i)$, respectively. }
    \label{fig:Cit_MTT_CRW_degree}
\end{figure}
%Figure~\ref{fig:Cit_MTT_CRW_degree} shows the dependence of $\mathrm{MTT}_i^{\mathrm{CRW}}$ on the out-degree and in-degree of the target node under CRW.
%As shown, the degree-based median $\mathrm{MTT}_i^{\mathrm{CRW}}$ decreases markedly with increasing $d_{\mathrm{in}}(i)$, whereas it exhibits no clear monotonic dependence on $d_{\mathrm{out}}(i)$ and fluctuates considerably at large out-degrees, because only a small number of nodes have these degree values.
%This indicates that, in the LSCC of the Cit-HepTh network, the trapping efficiency of CRW depends more strongly on the in-degree of the target node than on its out-degree.

Figs.~\ref{fig:Cit_MTT_CRW_degree} and \ref{fig:Cit_MTT_RSRW_degree} compare the degree dependence of the median $\mathrm{MTT}_i$ under CRW and RSRW within the Cit-HepTh LSCC, respectively.
For CRW, $\mathrm{MTT}_i^{\mathrm{CRW}}$ decreases sharply with increasing $d_{\mathrm{in}}(i)$, but exhibits no monotonic trend with $d_{\mathrm{out}}(i)$; fluctuations at large out-degrees stem from the limited number of nodes with such high degrees.
By contrast, RSRW displays a mirror-image behavior: $\mathrm{MTT}_i^{\mathrm{RSRW}}$ drops markedly with rising $d_{\mathrm{out}}(i)$, while showing no systematic dependence on $d_{\mathrm{in}}(i)$, with large-$d_{\mathrm{in}}$ fluctuations again attributable to sparse sampling of high-degree nodes.
Collectively, these results confirm that CRW trapping efficiency is governed predominantly by target-node in-degree, whereas RSRW efficiency is dictated by out-degree—a symmetry fully consistent with the theoretical equivalence between RSRW on $G$ and CRW on the reverse graph $G^r$ (Sec.~\ref{subsec:Comparison_RSRW_CRW}), and aligned with our earlier observations in synthetic Erdős–Rényi and scale-free networks.

\begin{figure}[tb]
    \centering
    \includegraphics[width=\linewidth]{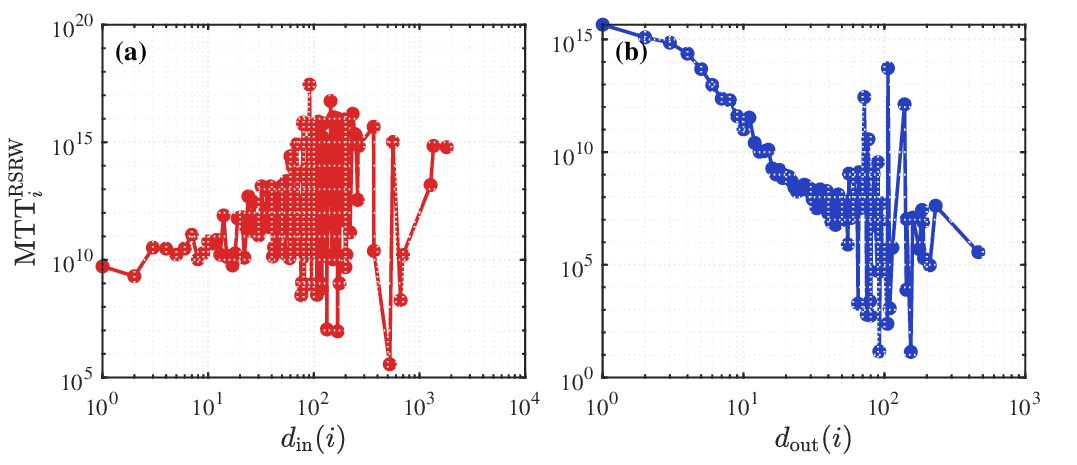}
    \caption{$\mathrm{MTT}_i^{\mathrm{RSRW}}$ is shown as a function of target-node degree for RSRW performed on the LSCC of the Cit-HepTh network. The red and blue curves in panels (a) and (b) show the degree-wise median $\mathrm{MTT}_i^{\mathrm{RSRW}}$ as functions of $d_{\mathrm{in}}(i)$ and $d_{\mathrm{out}}(i)$, respectively. }
    \label{fig:Cit_MTT_RSRW_degree}
\end{figure}
%Figure~\ref{fig:Cit_MTT_RSRW_degree} presents the corresponding results for RSRW.
%In contrast to CRW, the degree-based median $\mathrm{MTT}_i^{\mathrm{RSRW}}$ decreases markedly with increasing $d_{\mathrm{out}}(i)$, while showing no clear monotonic dependence on $d_{\mathrm{in}}(i)$.
%This indicates that the trapping efficiency of RSRW is governed primarily by the out-degree of the target node.

%Next,  we investigate whether their relative trapping efficiency can be characterized by the out-degree-to-in-degree ratio of the target node.

%To this end, for each target node $i$, we define the out-degree-to-in-degree ratio as

To further elucidate the relative exploration efficiency of CRW and RSRW within the largest strongly connected component (LSCC) of the Cit-HepTh network, we define the degree ratio
\begin{equation}
R_i = \frac{d_{\mathrm{out}}(i)}{d_{\mathrm{in}}(i)},
\label{eq:degree_ratio}
\end{equation}
and the relative mean trapping time
\begin{equation}
\tau_i = \frac{\mathrm{MTT}_i^{\mathrm{RSRW}}}{\mathrm{MTT}_i^{\mathrm{CRW}}}.
\label{eq:relative_mtt}
\end{equation}
We then analyze the relationship between $\tau_i$ and $R_i$. Results are shown in Fig.~\ref{fig:Cit_RiandTAUi}.
Here, $R_i$ quantifies the relative dominance of outgoing versus incoming connectivity for node $i$, while $\tau_i$ directly compares the trapping efficiencies of RSRW and CRW targeting the same node.
Specifically, $\tau_i < 1$ indicates that RSRW reaches node $i$ more efficiently than CRW, whereas $\tau_i > 1$ implies the opposite.

%Using the node-level MTT values obtained above for CRW and RSRW, we calculate the out-degree-to-in-degree ratio $R_i$ and the relative mean trapping time $\tau_i$ for each target node $i$ according to Eqs.~\eqref{eq:degree_ratio} and \eqref{eq:relative_mtt}, respectively.  Because both $R_i$ and $\tau_i$ span several orders of magnitude, a linear-scale representation would compress the majority of the data points into a small region owing to a limited number of large values, thereby obscuring the overall distribution and its underlying trend.
%We therefore display the relationship between $R_i$ and $\tau_i$ on logarithmic axes, together with a fitted trend line to indicate their overall dependence.

\begin{figure}[tb]
    \centering
    \includegraphics[width=0.75\columnwidth]{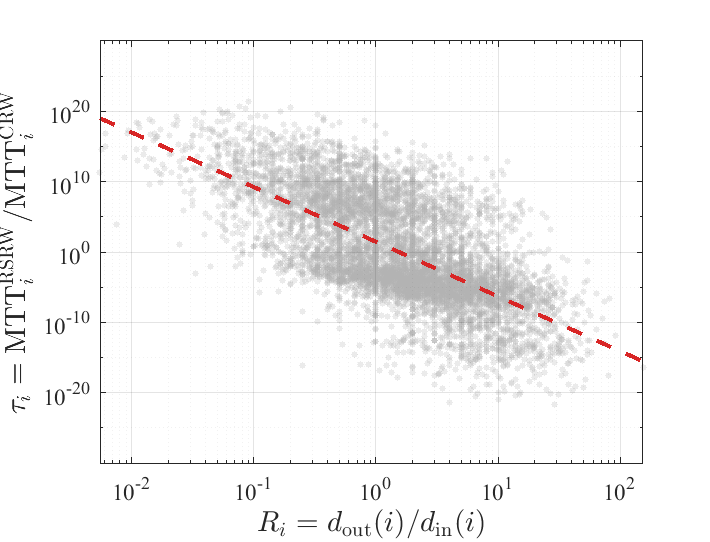}
    \caption{Relationship between the relative mean trapping time $\tau_i$ and the out-degree-to-in-degree ratio $R_i$ for target nodes in the LSCC of the Cit-HepTh network. The gray points represent individual target nodes, and the red dashed line is the fitted line for the $R_i$--$\tau_i$ data. Both axes are logarithmic.}
    \label{fig:Cit_RiandTAUi}
\end{figure}
As shown in Fig.~\ref{fig:Cit_RiandTAUi}, $\tau_i$ exhibits an overall decreasing trend with increasing $R_i$. This relationship converts the separate degree dependences of the two walks into a direct comparison of their exploration  efficiencies, suggesting that in-degree-dominated target nodes are statistically more favorable to CRW, whereas out-degree-dominated target nodes are more favorable to RSRW. Since review papers typically exhibit high out-degrees while hot papers are characterized by high in-degrees, our analysis indicates that CRW is more effective at identifying hot papers, whereas RSRW locates review papers more rapidly.
%由于综述论文具有较高的出度，热点论文具有较高的入度。 上面的分析结论表明，CRW更易于发现热点论文，而RSRW能更快的发现综述论文。

%In summary, the degree-dependent trapping behavior observed in the synthetic networks also persists in the real-world Cit-HepTh network: the MTT of CRW decreases mainly with the in-degree of the target node, whereas that of RSRW decreases mainly with its out-degree. Moreover, the overall decrease in $\tau_i$ with increasing $R_i$ converts these separate degree dependences into a direct comparison between the two walks. This result indicates that, as the outgoing connectivity of a target node becomes increasingly dominant over its incoming connectivity, the relative trapping advantage tends to shift from CRW to RSRW.

\subsection{Visit preferences of CRW and RSRW: A degree-dependent analysis}%{Finite-step exploration bias in the citation network}
\label{subsec:CAW_RSRW_preferences}
In this subsection, we perform finite-step computer simulations of CRW and RSRW on the LSCC.
By recording the node sequences (trajectories) at each step, we analyze and compare the in- and out-degrees of the nodes visited by the two mechanisms at identical trajectory positions.
This analysis explicitly reveals their distinct visit preferences with respect to nodal in-degree and out-degree.

%The preceding subsection characterized CRW and RSRW from the perspective of target-dependent trapping efficiency. Such first-passage quantities describe how readily a prescribed paper can be reached, but they do not directly reveal the structural types of papers encountered as a walk evolves over a limited number of steps. We therefore turn to a trajectory-level perspective and examine the finite-step visitation preferences of the two mechanisms on the LSCC of the Cit-HepTh network, thereby clarifying how their different transition rules translate into distinct exploration behaviors.

To mitigate stochastic fluctuations from individual walk realizations and ensure statistical robustness, we adopt the following simulation protocol.
For each trial, we uniformly select a random starting node from the LSCC and simultaneously simulate CRW and RSRW trajectories originating from this common node, recording the resulting node visitation sequences of length $L=50$ for both walk types.
We repeat this procedure $n=10^{6}$ times, yielding $10^{6}$ independent node sequences of length $L=50$ for CRW and RSRW, respectively.
Let $X_t$ denote the node visited by the walker at step $t$.
We define $\langle d_{\mathrm{in}}(X_t) \rangle$ and $\langle d_{\mathrm{out}}(X_t) \rangle$ as the ensemble-averaged in-degree and out-degree of nodes visited at step $t$ over the $10^{6}$ realizations, respectively.
Plotting $\langle d_{\mathrm{in}}(X_t) \rangle$ and $\langle d_{\mathrm{out}}(X_t) \rangle$ against $t$ reveals distinct visitation preferences for the two walk types, as summarized in Fig.~\ref{fig:Cit_Edegree}.
Specifically, Fig.~\ref{fig:Cit_Edegree}(a) shows that $\langle d_{\mathrm{out}}(X_t) \rangle$ remains consistently higher for RSRW than for CRW throughout the simulation.
Conversely, Fig.~\ref{fig:Cit_Edegree}(b) demonstrates that $\langle d_{\mathrm{in}}(X_t) \rangle$ is systematically larger for CRW than for RSRW over the same finite-step horizon.
These results confirm that, during finite-step exploration, RSRW exhibits a marked preference for nodes with high out-degree, whereas CRW preferentially targets nodes with high in-degree.

%To estimate these expectations, we uniformly select $$10^{6}$$ starting nodes without replacement from papers outside the top $5\%$ in out-degree and having at least five outgoing citation links. From each starting node, $200$ independent trajectories of length $10$ are generated for both CRW and RSRW, using the same starting-node set. After each trajectory reaches step $t$, we record the out-degree and in-degree of the node occupied at that step. The corresponding expectations are then estimated by averaging these degree values over all trajectories generated from all starting nodes.

\begin{figure}[tb]
    \centering
    \includegraphics[width=\linewidth]{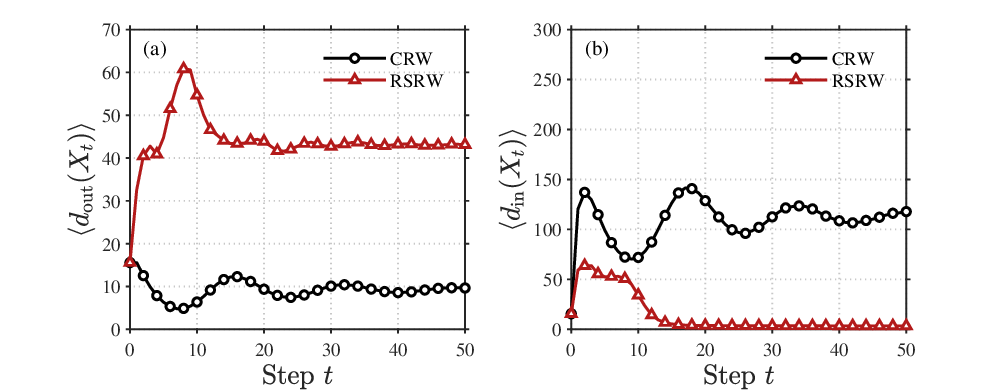}
    \caption{
        {Comparative analysis of nodal degree preferences: CRW vs.\ RSRW on the LSCC of the Cit-HepTh network.}
        {(a) $\langle d_{\mathrm{out}}(X_t) \rangle$ vs.\ step $t$, (b) $\langle d_{\mathrm{in}}(X_t) \rangle$ vs.\ step $t$.}
        Ensemble averages are computed over $10^{6}$ independent simulations.
    }
    \label{fig:Cit_Edegree}
\end{figure}

%As shown in Fig.~\ref{fig:Cit_Edegree}, panel~(a) shows that $\mathbb{E}[d_{\mathrm{out}}(X_t)]$ is consistently higher for RSRW than for CRW, whereas panel~(b) shows that $\mathbb{E}[d_{\mathrm{in}}(X_t)]$ is consistently higher for CRW than for RSRW over the considered finite steps. Accordingly, during finite-step exploration, RSRW preferentially visits nodes with larger out-degrees, whereas CRW preferentially visits nodes with larger in-degrees.
%our analysis indicates that CRW is more effective at identifying hot papers, whereas RSRW locates review papers more rapidly.

%In the context of citation networks, hot papers have high in-degree, review papers, methodological studies, or papers connecting different lines of research have higher out-degree. %上述结论表明，比较而言，RSRW更偏好于locates review papers

In the context of citation networks, these distinct visitation preferences provide a practical interpretation of the two walk mechanisms. CRW selects uniformly among the references of the current paper, whereas RSRW assigns nonuniform transition probabilities to the candidate references according to the stationary time-reversal construction based on the reverse citation graph. The consistently larger $\langle d_{\mathrm{out}}(X_t) \rangle$ observed for RSRW suggests that this transition bias tends to guide the walker toward reference-rich papers, which provide more outgoing citation pathways for subsequent exploration. Such papers may include review articles, methodological studies, or papers connecting different lines of research.
Accordingly, RSRW enables a distinct approach to literature exploration compared to CRW. Specifically, originating from a paper of interest, RSRW acts as a guided tracing strategy: it assists readers unfamiliar with a field in mapping its literature structure within a strictly limited number of citation-following steps.
%Accordingly, RSRW enables a distinct approach to literature exploration compared to CRW.
%Therefore, starting from a paper of interest, RSRW may serve as a guided literature-tracing strategy for readers unfamiliar with the field, helping them explore its literature structure within a limited number of citation-following steps.

}
\section{Conclusion} \label{sec:Conclusion}

Many real-world systems are inherently directed, necessitating random walk strategies that can effectively navigate asymmetric connectivity. Conventional approaches, however, are constrained to propagate downstream along the natural edge orientation, rendering the extraction of upstream topological statistics computationally prohibitive or practically infeasible. To bridge this gap, we introduced the \textit{reverse-graph stationary-probability-dependent random walk} (RSRW). By deriving transition probabilities from the stationary distribution of classical random walks (CRW) on the reverse graph, RSRW enables upstream exploration without physically reversing edges or altering the original network architecture.

We established rigorous theoretical proofs demonstrating that RSRW preserves the fundamental statistical properties of CRW on the reverse graph. Specifically, the stationary distribution, mean trapping time (MTT), and Kemeny's constant for RSRW on the original graph $G$ are mathematically identical to those of CRW on the reverse graph $G^r$. This equivalence was further validated through extensive numerical analysis on two canonical topologies and an arXiv citation network. Our comparative analyses revealed a striking duality in search efficiency: while CRW preferentially discovers nodes with high in-degree, RSRW exhibits superior efficacy in locating nodes with high out-degree. Consequently, RSRW redistributes stationary probability mass toward highly outgoing nodes, effectively inverting the topological sensitivity of conventional walks.

\acknowledgments

{
\noindent
The authors are grateful to Prof. Elena Agliari for  enlightening discussions  and acknowledge  the Basic and Applied Basic Research Foundation of Guangdong province (Grants No.2024A1515011700) .
}

\section*{DATA AVAILABILITY}
The data for the Cit-HepTh network are openly available in the Stanford Large Network Dataset Collection (SNAP) under the dataset name ``High-energy physics theory citation network (Cit-HepTh)'' at \url{https://snap.stanford.edu/data/cit-HepTh.html}. Other data that support the findings of this study are available
from the corresponding author upon reasonable request.

%The data that support the findings of this study are openly available in the Stanford Large Network Dataset Collection (SNAP) under the dataset name ``High-energy physics theory citation network (Cit-HepTh)'' at \url{https://snap.stanford.edu/data/cit-HepTh.html}.

\appendix

\section{Representative directed Erd\H{o}s--R\'{e}nyi  network and corresponding stationary distributions}
\label{appendix:der-stationary-distribution}

Here, we present a representative directed Erd\H{o}s--R\'{e}nyi network with $N=20$ and $p=0.15$, along with the corresponding stationary distributions for CRW on $G$, CRW on $G^r$, and RSRW on $G$. This example demonstrates that the stationary distribution of RSRW on $G$ differs from that of CRW on $G$, yet coincides with that of CRW on $G^r$.

\begin{figure}[!b]
    \centering
    \includegraphics[width=\columnwidth]{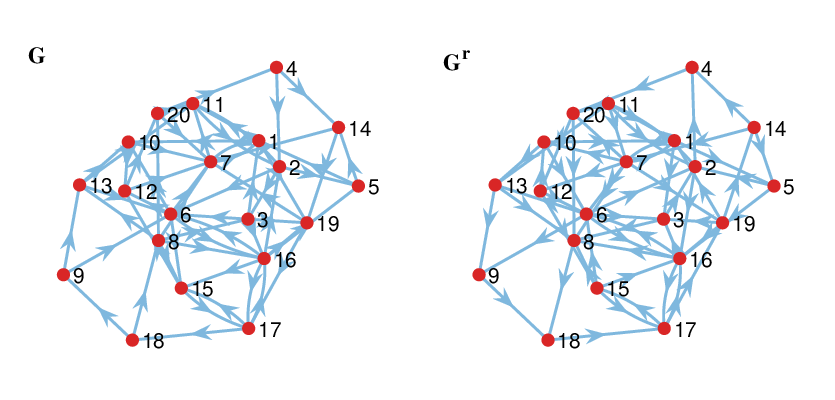}
    \caption{Structural comparison of a directed Erd\H{o}s--R\'{e}nyi network $G(N,p)$ with $N=20$ and $p=0.15$ (left) and its reverse graph $G^r$ (right). Both panels share the same node layout.}
    \label{fig:der-graph-compare-g-gr}
\end{figure}

Figure~\ref{fig:der-graph-compare-g-gr} illustrates the original graph $G$ and its reverse graph $G^r$.

\begin{figure}[!b]
    \centering
    \includegraphics[width=0.9\columnwidth]{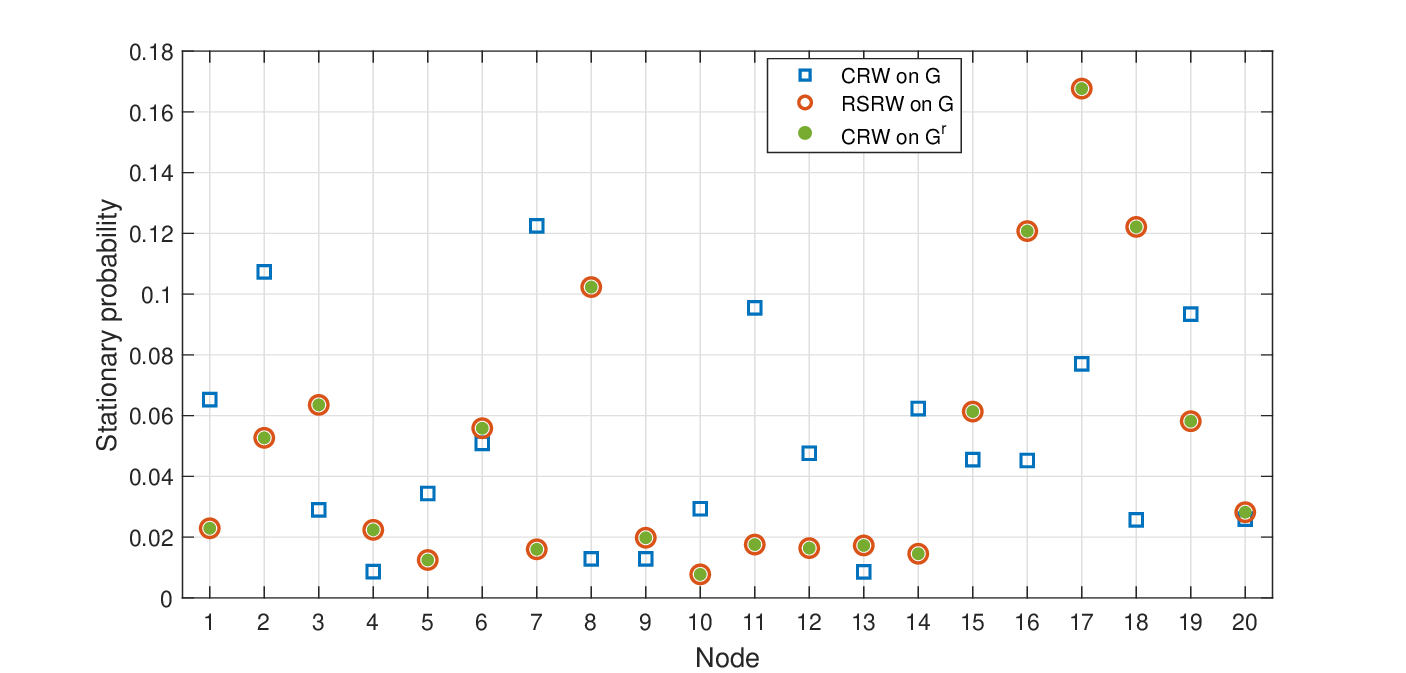}
    \caption{Scatter plot comparison of the three stationary probability distributions shown in Table~\ref{tab:der-stationary-distribution}.  Blue squares denote  stationary probabilities for CRW on $G$; green dots denote stationary probabilities for CRW on $G^r$; red circles denote stationary probabilities for RSRW on $G$.
    }%Stationary probabilities of the three random-walk strategies at each node. The calculations are performed on the directed graph $G$ shown in the left panel of Fig.~\ref{fig:der-graph-compare-g-gr}. Blue squares denote the classical random walk on the original graph, red circles denote the new random-walk strategy on the original graph, and green dots denote the classical random walk on the reverse graph.
    \label{fig:der-stationary-distribution-compare}
\end{figure}

Table~\ref{tab:der-stationary-distribution} lists the three stationary distributions for this network, while Fig.~\ref{fig:der-stationary-distribution-compare} presents the corresponding scatter plot comparison. We emphasize that for any initial distribution $\vec{p}(0)$, the limiting distribution $\lim_{t \to \infty} \vec{p}(0) P^t$ coincides with the stationary distribution reported in Table~\ref{tab:der-stationary-distribution}, where $P$ is the respective transition matrix.

\begin{table}[!t]
\centering
% 备选版本：清晰罗列
\caption{Node degrees and stationary probability distributions for the network shown in Fig.~\ref{fig:der-graph-compare-g-gr}. Column labels: Node,  $d_i^{\mathrm{out}}$, and $d_i^{\mathrm{in}}$ denote the node index, out-degree, and in-degree for graph $G$, respectively; $\pi_i^{CRW}$, $\pi_i^{RSRW}$, and $\pi_i^{RG}$ denote the stationary distributions of the CRW on $G$, the RSRW on $G$, and the CRW on $G^r$, respectively.}

\label{tab:der-stationary-distribution}
\small
\begin{tabular*}{\columnwidth}{@{\extracolsep{\fill}}cccccc}
\hline\hline
Node & $d_i^{\mathrm{out}}$ & $d_i^{\mathrm{in}}$ & $\pi_i^{CRW}$ & $\pi_i^{RSRW}$ & $\pi_i^{RG}$ \\
\hline
1 & 4 & 3 & 0.065269 & 0.022929 & 0.022929 \\
2 & 4 & 3 & 0.107361 & 0.052709 & 0.052709 \\
3 & 4 & 2 & 0.028980 & 0.063553 & 0.063553 \\
4 & 2 & 1 & 0.008651 & 0.022418 & 0.022418 \\
5 & 2 & 2 & 0.034381 & 0.012491 & 0.012491 \\
6 & 5 & 5 & 0.050899 & 0.055867 & 0.055867 \\
7 & 3 & 7 & 0.122501 & 0.016005 & 0.016005 \\
8 & 6 & 1 & 0.012839 & 0.102339 & 0.102339 \\
9 & 2 & 1 & 0.012839 & 0.019809 & 0.019809 \\
10 & 2 & 3 & 0.029350 & 0.007752 & 0.007752 \\
11 & 1 & 5 & 0.095495 & 0.017570 & 0.017570 \\
12 & 2 & 3 & 0.047624 & 0.016396 & 0.016396 \\
13 & 3 & 2 & 0.008560 & 0.017271 & 0.017271 \\
14 & 1 & 3 & 0.062349 & 0.014545 & 0.014545 \\
15 & 2 & 4 & 0.045539 & 0.061358 & 0.061358 \\
16 & 6 & 4 & 0.045243 & 0.120766 & 0.120766 \\
17 & 3 & 3 & 0.077036 & 0.167678 & 0.167678 \\
18 & 2 & 1 & 0.025679 & 0.122147 & 0.122147 \\
19 & 2 & 4 & 0.093452 & 0.058179 & 0.058179 \\
20 & 3 & 2 & 0.025952 & 0.028218 & 0.028218 \\
\hline\hline
\end{tabular*}
\end{table}

As shown in Table~\ref{tab:der-stationary-distribution} and Fig.~\ref{fig:der-stationary-distribution-compare}, the stationary distribution generated by RSRW on the original directed graph coincides with that of CRW on the reverse graph, while differing substantially from the stationary distribution of CRW on the original graph. This observation is fully consistent with the analytical results obtained in the previous section, thereby providing further numerical validation of the theoretical derivation.

\section{Derivation of the Asymptotic Degree Distribution}
\label{appendix:Asymptotic-Degree-Distribution}
In this appendix, we use mean-field theory to derive the out-degree and in-degree distributions of the model proposed in Sec.~\ref{subsec:DSF-network}.

Let $N(t)$ and $M(t)$ denote the number of vertices and directed edges, respectively, in the network at time $t$. For the modified model, they satisfy
\begin{equation}
N(t)\sim \mu t,\qquad M(t)\sim (1+\mu)t .
\label{appeq:NM-asymptotic}
\end{equation}

The quantity $d_i^{\mathrm{in}}(t)+\delta_{\mathrm{in}}$ gives the attractiveness of vertex $i$ in the in-degree preferential-attachment process.
So, the corresponding total attractiveness is
\begin{equation}
\begin{aligned}
\sum_{i\in V(t)}
\left[d_i^{\mathrm{in}}(t)+\delta_{\mathrm{in}}\right]
&=
M(t)+\delta_{\mathrm{in}}N(t)
\\
&\sim
\left[1+\mu+\mu\delta_{\mathrm{in}}\right]t
\\
&\overset{\triangle}{=}
A_{\mathrm{in}}t ,
\end{aligned}
\label{appeq:in-total-attractiveness}
\end{equation}
where $A_{\mathrm{in}}=1+\mu+\mu\delta_{\mathrm{in}}$.

Let $X_k^{\mathrm{in}}(t)$ denote the number of vertices with in-degree $k$ at time $t$. Then the mean-field master equation for $X_k^{\mathrm{in}}(t)$ is
\begin{equation}
\begin{aligned}
&\mathbb{E}\!\left[
X_k^{\mathrm{in}}(t+1)-X_k^{\mathrm{in}}(t)
\right]
\\
&\quad =
\mu \mathbf{1}_{\{k=1\}}
\\
&\qquad
+
\frac{
(k-1+\delta_{\mathrm{in}})X_{k-1}^{\mathrm{in}}(t)
-
(k+\delta_{\mathrm{in}})X_k^{\mathrm{in}}(t)
}{
M(t)+\delta_{\mathrm{in}}N(t)
}.
\end{aligned}
\label{appeq:in-master-equation}
\end{equation}
Here, the first term represents the contribution of a newly introduced vertex, whose in-degree is one. The second term describes the gain of vertices of in-degree $k$ from vertices of in-degree $k-1$, while the third term describes the loss of vertices of in-degree $k$ when they are selected and their in-degree increases to $k+1$.

Assume that, as $t\to\infty$,
\begin{equation}
\frac{X_k^{\mathrm{in}}(t)}{t}
\sim
p_k^{\mathrm{in}} .
\label{appeq:X-in-asymptotic}
\end{equation}
The quantity $p_k^{\mathrm{in}}$ is a function of the degree $k$ and describes the asymptotic growth rate of the number of vertices with in-degree $k$.

Under the mean-field assumption in Eq.~\eqref{appeq:X-in-asymptotic}, $X_k^{\mathrm{in}}(t)$ grows linearly with $t$, namely,
$X_k^{\mathrm{in}}(t)\sim p_k^{\mathrm{in}}t$. Hence,
\[
X_k^{\mathrm{in}}(t+1)-X_k^{\mathrm{in}}(t)
\sim
p_k^{\mathrm{in}}(t+1)-p_k^{\mathrm{in}}t
=
p_k^{\mathrm{in}} .
\]
Thus, in the mean-field approximation, the left-hand side of Eq.~\eqref{appeq:in-master-equation} is replaced by $p_k^{\mathrm{in}}$.
Substituting Eqs.~\eqref{appeq:in-total-attractiveness} and~\eqref{appeq:X-in-asymptotic} into Eq.~\eqref{appeq:in-master-equation} gives
\begin{equation}
p_k^{\mathrm{in}}
=
\frac{
(k-1+\delta_{\mathrm{in}})p_{k-1}^{\mathrm{in}}
-
(k+\delta_{\mathrm{in}})p_k^{\mathrm{in}}
}{
A_{\mathrm{in}}
}
+
\mu \mathbf{1}_{\{k=1\}} .
\label{appeq:p-in-balance}
\end{equation}
Equivalently,
\begin{equation}
\left(
k+\delta_{\mathrm{in}}+A_{\mathrm{in}}
\right)
p_k^{\mathrm{in}}
=
(k-1+\delta_{\mathrm{in}})p_{k-1}^{\mathrm{in}}
+
A_{\mathrm{in}}\mu \mathbf{1}_{\{k=1\}} .
\label{appeq:p-in-recursion-general}
\end{equation}
For $k>1$, the term $\mu \mathbf{1}_{\{k=1\}}$ vanishes, and therefore
\begin{equation}
p_k^{\mathrm{in}}
=
\frac{
k-1+\delta_{\mathrm{in}}
}{
k+\delta_{\mathrm{in}}+A_{\mathrm{in}}
}
p_{k-1}^{\mathrm{in}} .
\label{appeq:p-in-recursion}
\end{equation}
Iterating Eq.~\eqref{appeq:p-in-recursion}, we obtain
\begin{equation}
p_k^{\mathrm{in}}
=
p_1^{\mathrm{in}}
\prod_{r=2}^{k}
\frac{
r-1+\delta_{\mathrm{in}}
}{
r+\delta_{\mathrm{in}}+A_{\mathrm{in}}
}.
\label{appeq:p-in-product}
\end{equation}

Using the Gamma function identity $\Gamma(x+1)=x\Gamma(x)$, the numerator of Eq.~\eqref{appeq:p-in-product} can be written as
\begin{equation}
\prod_{r=2}^{k}
\left(
r-1+\delta_{\mathrm{in}}
\right)
=
\prod_{s=1}^{k-1}
\left(
s+\delta_{\mathrm{in}}
\right)
=
\frac{
\Gamma(k+\delta_{\mathrm{in}})
}{
\Gamma(1+\delta_{\mathrm{in}})
}.
\label{appeq:gamma-numerator}
\end{equation}
Similarly, the denominator is
\begin{equation}
\prod_{r=2}^{k}
\left(
r+\delta_{\mathrm{in}}+A_{\mathrm{in}}
\right)
=
\frac{
\Gamma(k+1+\delta_{\mathrm{in}}+A_{\mathrm{in}})
}{
\Gamma(2+\delta_{\mathrm{in}}+A_{\mathrm{in}})
}.
\label{appeq:gamma-denominator}
\end{equation}
Hence,
\begin{equation}
p_k^{\mathrm{in}}
=
p_1^{\mathrm{in}}
\frac{
\Gamma(k+\delta_{\mathrm{in}})
\Gamma(2+\delta_{\mathrm{in}}+A_{\mathrm{in}})
}{
\Gamma(1+\delta_{\mathrm{in}})
\Gamma(k+1+\delta_{\mathrm{in}}+A_{\mathrm{in}})
}.
\label{appeq:p-in-gamma}
\end{equation}
Using the following asymptotic relation for the Gamma function $\Gamma(\cdot)$, we have
\begin{equation}
\frac{\Gamma(k+a)}{\Gamma(k+b)}
\sim
k^{a-b},
\qquad k\to\infty .
\label{appeq:gamma-asymptotic-relation}
\end{equation}
Applying Eq.~\eqref{appeq:gamma-asymptotic-relation}, we obtain
\begin{equation}
p_k^{\mathrm{in}}
\sim
p_1^{\mathrm{in}}
\frac{
\Gamma(2+\delta_{\mathrm{in}}+A_{\mathrm{in}})
}{
\Gamma(1+\delta_{\mathrm{in}})
}
k^{-(1+A_{\mathrm{in}})} .
\label{appeq:p-in-asymptotic}
\end{equation}

Therefore,
\begin{equation}
p_k^{\mathrm{in}}
\sim
k^{-\left(1+A_{\mathrm{in}}\right)}
=
k^{-\left[2+\mu(1+\delta_{\mathrm{in}})\right]} .
\label{appeq:p-in-power-law}
\end{equation}
So, the in-degree distribution has the asymptotic form
\begin{equation}
P_{\mathrm{in}}(k)
\sim
k^{-\left[2+\mu(1+\delta_{\mathrm{in}})\right]} .
\label{appeq:in-degree-exponent}
\end{equation}

Similarly, the asymptotic form of the out-degree distribution is obtained as
\begin{equation}
P_{\mathrm{out}}(k)
\sim
k^{-\left[2+\mu(1+\delta_{\mathrm{out}})\right]} .
\label{appeq:out-degree-exponent}
\end{equation}

\section{Representative directed scale-free  network and corresponding stationary distributions}
\label{appendix:DSF-stationary-distribution}

Here, we present a representative directed scale-free network with $N=20$, $\mu=0.6$, and $\delta_{\mathrm{in}}=\delta_{\mathrm{out}}=1$, along with the corresponding stationary distributions for CRW on $G$, CRW on $G^r$, and RSRW on $G$. This example demonstrates that the stationary distribution of RSRW on $G$ differs from that of CRW on $G$, yet coincides with that of CRW on $G^r$.

\begin{figure}[tb]
    \centering
    \includegraphics[width=\columnwidth]{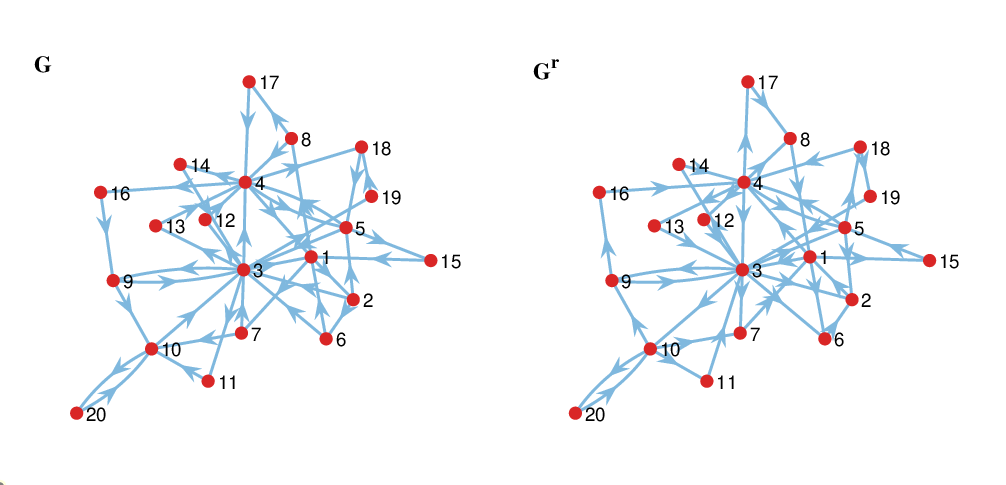}
    \caption{Structural comparison of a directed scale-free network $G$ (left) and its reverse graph $G^r$ (right) with $N=20$, $\mu=0.6$, and $\delta_{\mathrm{in}}=\delta_{\mathrm{out}}=1$. Nodes and edges are arranged identically to highlight directional differences.}
    \label{fig:DSF_graph_compare_G_Gr}
\end{figure}

We generate $G$ and $G^r$ following the construction in Sec.~\ref{subsec:DSF-network}. Figure~\ref{fig:DSF_graph_compare_G_Gr} visualizes their structural difference.

\begin{figure}[b]
    \centering
    \includegraphics[width=0.90\columnwidth]{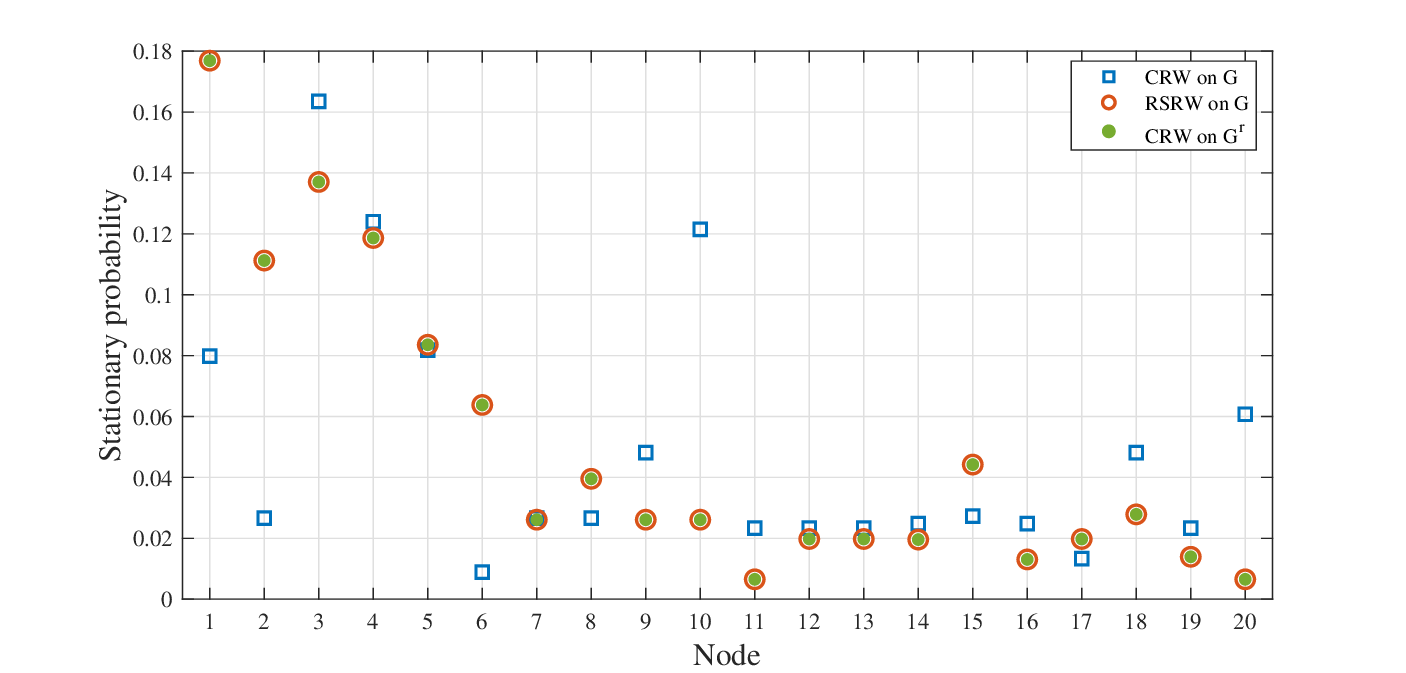}
    \caption{Scatter plot comparing the three stationary distributions (CRW on $G$, CRW on $G^r$, and RSRW on $G$) for this instance is presented in Fig.~\ref{fig:DSF_graph_compare_G_Gr}. Blue squares denote  stationary probabilities for CRW on $G$; green dots denote stationary probabilities for CRW on $G^r$; red circles denote stationary probabilities for RSRW on $G$.}
    %\caption{Stationary probabilities of the three random-walk strategies at each node. The calculations are performed on the directed graph $G$ shown in the left panel of Fig.~\ref{fig:DSF_graph_compare_G_Gr}. Blue squares denote the classical random walk on the original graph, red circles denote the new random-walk strategy on the original graph, and green dots denote the classical random walk on the reverse graph.}
    \label{fig:DSF_stationary_distribution_compare}
\end{figure}

The transition matrices for the three walks are constructed via Eqs.~\eqref{CRW-p}, \eqref{eq:RG-p-AD}, and \eqref{RSRW-p}, respectively. Solving the stationary condition \eqref{stationary-equ} yields the distributions listed in Table~\ref{tab:dsf-stationary-distribution}. Figure~\ref{fig:DSF_stationary_distribution_compare} provides a scatter plot comparison, confirming that the stationary distribution of RSRW on $G$ coincides with that of CRW on $G^r$ while differing markedly from that of CRW on $G$. These numerical results align fully with the analytical equivalence established above.

We emphasize that $\vec{\pi}$ is the unique limiting distribution: for any initial $\vec{p}(0)$, $\lim_{t \to \infty} \vec{p}(0) P^{t} = \vec{\pi}$, where $P$ denotes the respective transition matrix.

\begin{table}[!tb]
\centering
\caption{Node degrees and stationary probability distributions for the network shown in Fig.~\ref{fig:DSF_graph_compare_G_Gr}. Column labels: Node,  $d_i^{\mathrm{out}}$, and $d_i^{\mathrm{in}}$ denote the node index, out-degree, and in-degree for graph $G$, respectively; $\pi_i^{CRW}$, $\pi_i^{RSRW}$, and $\pi_i^{RG}$ denote the stationary probabilities of the CRW on $G$, the RSRW on $G$, and the CRW on $G^r$, respectively.}
\label{tab:dsf-stationary-distribution}
\small
\begin{tabular*}{\columnwidth}{@{\extracolsep{\fill}}cccccc}
\hline\hline
Node & $d_i^{\mathrm{out}}$ & $d_i^{\mathrm{in}}$ & $\pi_i^{CRW}$ & $\pi_i^{RSRW}$ & $\pi_i^{RG}$ \\
\hline
1 & 3 & 4 & 0.079859 & 0.176877 & 0.176877 \\
2 & 3 & 1 & 0.026620 & 0.111230 & 0.111230 \\
3 & 7 & 7 & 0.163502 & 0.137041 & 0.137041 \\
4 & 5 & 6 & 0.123964 & 0.118633 & 0.118633 \\
5 & 3 & 3 & 0.081816 & 0.083569 & 0.083569 \\
6 & 2 & 1 & 0.008873 & 0.063797 & 0.063797 \\
7 & 2 & 1 & 0.026620 & 0.026103 & 0.026103 \\
8 & 2 & 1 & 0.026620 & 0.039544 & 0.039544 \\
9 & 2 & 2 & 0.048150 & 0.026103 & 0.026103 \\
10 & 2 & 4 & 0.121485 & 0.026103 & 0.026103 \\
11 & 1 & 1 & 0.023357 & 0.006526 & 0.006526 \\
12 & 1 & 1 & 0.023357 & 0.019772 & 0.019772 \\
13 & 1 & 1 & 0.023357 & 0.019772 & 0.019772 \\
14 & 1 & 1 & 0.024793 & 0.019577 & 0.019577 \\
15 & 1 & 1 & 0.027272 & 0.044219 & 0.044219 \\
16 & 1 & 1 & 0.024793 & 0.013052 & 0.013052 \\
17 & 1 & 1 & 0.013310 & 0.019772 & 0.019772 \\
18 & 1 & 2 & 0.048150 & 0.027856 & 0.027856 \\
19 & 1 & 1 & 0.023357 & 0.013928 & 0.013928 \\
20 & 1 & 1 & 0.060742 & 0.006526 & 0.006526 \\
\hline\hline
\end{tabular*}
\end{table}

%%%%%%%%%%%%%%%
%%% APPENDIX A%%%%
%%%%%%%%%%%%%%%

%\section{Derivation of Eqs.~(\ref{Wja}) and (\ref{EQ_Sigma})}
%\label{Sec:WB}

%\bibliographystyle{apsrev}%{plain}
%\bibliography{GZPref}

%\section*{References}
%\begin{thebibliography}{0}
%[31]Chandra, A. K., Raghavan, P., Ruzzo, W. L., Smolensky, R., and Tiwari, P. (1989). The electrical resistance of a graph captures its commute and cover times,Proceedings of the 21st Annual ACM Symposium on Theory of Computing, May. ACM Press, Seattle.
%
%\end{thebibliography}
\end{document}